\documentclass[english,journal=jpccck]{achemso}
\pdfoutput=1
\usepackage{textcomp}
\usepackage[utf8]{inputenc}
\usepackage{color}
\usepackage{array}
\usepackage{booktabs}
\usepackage{amsmath}
\usepackage{graphicx}

\makeatletter

\title{DFT and Model Hamiltonian Study of Optoelectronic Properties of Some
Low-Symmetry Graphene Quantum Dots}

\author{Samayita Das }

\email{samayitadas5@gmail.com}

\affiliation{Department of Physics, Indian Institute of Technology Bombay, Powai,
Mumbai 400076, India}

\author{Alok Shukla}

\email{shukla@iitb.ac.in}

\affiliation{Department of Physics, Indian Institute of Technology Bombay, Powai,
Mumbai 400076, India}

\providecommand{\tabularnewline}{\\}

\@ifundefined{date}{}{\date{}}
\SectionNumbersOn

\makeatother

\usepackage{babel}
\begin{document}
\begin{abstract}
We have studied the electronic and optical properties of three low-symmetry
graphene quantum dots (GQDs), with the point-group symmetries $C_{2v}$,
and $C_{2h}$.  For the calculations of linear optical absorption
spectra , we employed both the first-principles time-dependent density-functional
theory (TDDFT), and the electron-correlated Pariser-Parr-Pople (PPP)
model coupled with the configuration-interaction (CI) approach. In
the PPP-CI approach, calculations were performed using both screened
and standard parameters, along with efficiently incorporating electron
correlation effects using multi-reference singles-doubles configuration-interaction
for both ground and excited states. We assume that the GQDs are saturated
by hydrogen atoms at the edges, making them effectively polycyclic
aromatic hydrocarbons (PAHs) dibenzo{[}bc,ef{]}coronene (also known
as benzo(1,14)bisanthene, C\textsubscript{30}H\textsubscript{14}),
and two isomeric compounds, dinaphtho{[}8,1,2abc;2´,1´,8´klm{]}coronene
and dinaphtho{[}8,1,2abc;2´,1´,8´jkl{]}coronene with the chemical
formula C\textsubscript{36}H\textsubscript{16}. The two isomers
have different point group symmetries, therefore, this study will
also help us understand the influence of symmetry on the optical properties.
A common feature of the absorption spectra of the three GQDs is that
the first peak representing the optical gap is of low to moderate
intensity, while the intense peaks appear at higher energies. For
each GQD, PPP model calculations performed with the screened parameters
agree well with the experimental results of the corresponding PAH,
and also with the  TDDFT calculations. To further quantify the influence
of electron-correlation effects, we also computed the singlet-triplet
gap (spin gap) of the three GQDs, and we found them to be significant.
\end{abstract}

\section{Introduction}

The electronic structure of an isolated monolayer of graphite, called
graphene in the modern times, was studied long ago within the tight-binding
model by Wallace, who demonstrated the linear dispersion of the system
in the Brillouin zone, and its Dirac cone like structure \cite{prwallace-graphite-1947}.
Indeed, within a few years of its synthesis \cite{graphene-synthesis-2004},
graphene became one of the most studied quantum material in general,
and Dirac material, in particular \cite{graphene-review-castro-neto-rmp-2009}.
Graphene is noteworthy for its remarkable transport properties because
of the massless charge carriers and a very high mobility \cite{graphene-synthesis-2004}.
However, given the vanishing band gap of graphene, its applications
in the field of optoelectronics are limited. Therefore, there has
been significant research in the field to introduce a gap in graphene
by chemical modifications such as oxidation, hydrogenation, and heteroatom
doping \cite{wu2010light,son2016hydrogenated,denis2010band}. Another
way to introduce a gap is by achieving quantum confinement by breaking
the translation symmetry of graphene and considering its finite-sized
fragments referred to as nano-flakes, nano-disks, and graphene quantum
dots (GQDs) \cite{li2010colloidal,huang2011graphene,bacon2014graphene}.
The problem with the finite-sized fragments is that they will undergo
edge reconstruction to saturate the dangling bonds, and the resulting
structure could be highly asymmetric and nonplanar. Therefore, in
order to keep a fragment planar, the dangling bonds on its edges need
to be passivated. If the interior of a fragment contains multiple
aromatic rings, and its dangling bonds are saturated by the hydrogen
atoms, the resultant structure will continue to be fully planar, and
chemically speaking it will be nothing but a polycyclic aromatic hydrocarbon
(PAH) and this has been realized by authors in the past who argued
that suitably chosen PAHs can be used as models of mono- and multilayer
graphite fragments \cite{zerner-group-pah-apj-1991}. Therefore, in
this work we study the electronic structure and optical properties
of three PAHs of somewhat lower point-group symmetries as models of
corresponding planar GQDs.

Because of their strictly planar geometry and aromatic character,
PAHs are $\pi$-conjugated organic molecules in which the energetically
low-lying $\sigma$-bonds hold the atomic skeleton together, while
the orbitals close to the Fermi level are of the $\pi$ type. PAHs
are abundant in nature because they are not only found in fossil fuels
on earth, but also in the interstellar dust and are believed to be
responsible for the infrared emission bands dominating the spectra
emerging from galactic and extra-galactic sources\cite{annurev:/content/journals/10.1146/annurev.astro.46.060407.145211}
as well as for the near infrared to visible absorption bands called
diffuse interstellar bands \cite{diffuse-interstellar-band-Geballe_2016}.
Because of this reason, the electronic structure, vibrational and
optical properties of PAHs have been of great interest not only in
physics and chemistry, but also in astronomy and astrophysics \cite{zerner-group-pah-apj-1991,diffuse-interstellar-band-Geballe_2016,annurev:/content/journals/10.1146/annurev.astro.46.060407.145211}.
Given the itinerant character of the $\pi$ electrons, their response
to externally applied electromagnetic field is strong, leading to
interesting linear and nonlinear optical properties of PAHs. As a
result of which, PAHs (or GQDs), have tremendous potential applications
in optoelectronic devices \cite{lussem2016doped,ma2010interface,facchetti2011pi,guo2017fabrication,liu2018all,fetzer2007chemistry,friend1999electroluminescence,burroughes1990light}.
Due to their interesting photophysical properties, PAHs can also be
used as solute polarity probes \cite{acree1990polycyclic}. In addition,
it is a well-known fact that many PAHs are highly carcinogenic, therefore,
it is important to study their electronic structure and related properties
so as to understand their interaction with living tissues \cite{zerner-group-pah-apj-1991,harvey1991polycyclic}.

In this work we have studied the optical and electronic properties
of three polycyclic aromatic hydrocarbons (PAHs), namely, dibenzo{[}bc,ef{]}coronene
which is also known as 1.14-benzobisanthene (C\textsubscript{30}H\textsubscript{14}),
dinaphtho{[}8,1,2abc;2´,1´,8´klm{]}coronene (C\textsubscript{36}H\textsubscript{16})
and dinaphtho{[}8,1,2abc;2´,1´,8´jkl{]}coronene (C\textsubscript{36}H\textsubscript{16}).
For the purpose we have used both the first-principles density-functional
theory (DFT) based approach as well as the effective $\pi$-electron
methodology based on Pariser-Parr-Pople (PPP) Hamiltonian \cite{pople1953electron,pariser1953semi}.
Among the three different PAHs, the last two molecules with 11 aromatic
rings have the same chemical formula C\textsubscript{36}H\textsubscript{16}
as they are isomers of each other, but their optical absorption spectra
are quite different because of different point-group symmetries. By
comparing we find very good agreement between our calculated absorption
spectra and the experimental ones. We have compared our results for
1.14-benzobisanthene (C\textsubscript{30}H\textsubscript{14}) with
the reported data of the important peaks in the experimental absorption
spectrum measured by Clar and Schmidt\cite{clar1977correlations}.
For the other two isomeric GQDs (C\textsubscript{36}H\textsubscript{16}),
we have compared the theoretically computed absorption spectra with
the measured ones reported by Fetzer \cite{fetzer2000large}, and
Bryson \emph{et al}. \cite{BRYSON20111980}. 

The remainder of the article is organized as follows. In section \ref{sec:Computational-Details}
we describe our computational methodology, while in the section \ref{sec:Results-and-Discussion}
we present and discuss our results. Finally, in section \ref{sec:Conclusions}
we summarize our conclusions.

\section{Computational Methods}

\label{sec:Computational-Details}

In principle it will be ideal to perform first-principles calculations
using a wave-function-based approach such as coupled-cluster, high-order
configuration interaction, etc., on the molecules considered in this
work. However, given the size of the systems under consideration,
it is extremely computationally intensive to do so, and to the best
of our knowledge, no such calculations on these systems have been
performed earlier. Within the first-principles framework, calculations
are performed routinely using the DFT-based methodologies, while accurate
wave-function-based calculations can be performed employing approximate
Hamiltonians such as the PPP model. Next, we briefly describe the
DFT-based first-principles approach along with the PPP-model based
configuration interaction (CI) approach, that we employed to study
the optical properties of the H-saturated GQDs.

\subsection{First-Principles Approach}

The first-principles approach used by us is based on the DFT methodology
as implemented in the Gaussian-basis-functions based Gaussian16 package
\cite{g16}. In our calculations, for all the three molecules, we
chose the valence double zeta 6-31++G(d,p) basis set which includes
polarization and diffuse functions \cite{hehre1972self,hariharan1973influence},
coupled with the hybrid exchange-correlation functional B3LYP \cite{lee1988development,becke1992density,becke1993density}. 

Convergence criteria of $10^{-8}$ Hartree was set to self consistently
solve Kohn-Sham equations \cite{kohn1965self}. The geometry of a
molecule was considered converged only after the maximum force on
an atom, average (RMS) force, maximum displacement, and the average
(RMS) displacement were less than 0.00045 Hartree/Bohr, 0.00030 Hartree/Bohr,
0.0018 Bohr, and 0.0012 Bohr, respectively. Even though all the three
GQDs considered in this work are well known PAHs, we still confirmed
the stability of their computed structures by performing vibrational
frequency analysis using the force constants computed in the DFT calculations.
For none of the molecules any imaginary frequencies were found indicating
that all the three PAHs considered in this work are dynamically stable.
For calculation of the total and partial density of states multiwfn
software is used \cite{lu2012multiwfn}. 

In order to compute the optical absorption spectrum of a GQD, we need
to know the excitation energies of its excited states along with their
transition dipole couplings to the ground state. For the purpose,
we employed the time--dependent density-functional-theory (TDDFT)\cite{runge-gross-tddft-1984,marques2004time}
approach as implemented in Gaussian16\cite{g16}, along with the same
exchange correlation functional (B3LYP) which was used for the ground
state calculations. Specifically, the Gaussian16 implementation\cite{g16}
uses an adiabatic frequency-space based approach\cite{BAUERNSCHMITT1996454,casida1998molecular,stratmann1998efficient,VANCAILLIE1999249,VANCAILLIE2000159,furche2002adiabatic,scalmani2006geometries},
and in our calculations we utilized an ``ultrafine'' integration
grid consisting of 99 radial shells and 590 angular points per shell,
and an energy convergence threshold of $10^{-6}$ eV. We did not consider
geometry relaxation for the excited states and computed 40 vertically
excited states using the TDDFT approach for all the three molecules.
The TDDFT method is similar to the SCI (single-CI) approach as both
consider only 1 particle\textminus 1 hole excitations, as a result
of which it is accurate for singly-excited optical states. 

Previously published works on PAHs used similar functional and basis
set combinations, and reported good results on optical properties
\cite{hirata2003time,bhattacharyya2020pariser,C9CP01038F,parac2003tddft,malloci2004electronic,malloci2007line}.
However, for the purpose of benchmarking, we performed calculations
using two other functionals, namely HSE06 and PBE on dibenzo{[}bc,ef{]}coronene
(C\textsubscript{30}H\textsubscript{14}), and the results are summarized
in Table S10 of Supporting Information. From the table it is obvious
that HSE06 results on various peak locations are in excellent agreement
with the corresponding B3LYP results, while PBE functional, as expected,
underestimates the optical gap, leading to results in disagreement
with experiments. Therefore, based on these results, as well as earlier
studies, we finally adopted the combination B3LYP/6-31++G(d,p), which
shows very good comparison with the experiment. 

\subsection{PPP Model Based Approach}

In addition to the first-principles approach described above, we have
also used an effective $\pi$-electron approach to compute the low-lying
excited states and the optical absorption spectra of the molecules
considered in this work. Next, we briefly describe the underlying
model Hamiltonian, its parameterization, and the electron-correlated
configuration interaction approach employed to compute the optical
spectra of the considered GQDs. 

\subsubsection{The Hamiltonian and its Parameters}

In $\pi$-conjugated materials, $\sigma$ and $\pi$ electrons are
well separated in energies, with the $\pi$ electrons being itinerant
and close to the Fermi level, while the $\sigma$ electrons being
highly localized with the energies away from the Fermi level. Based
on this concept of $\sigma-\pi$ separation, Pariser, Parr, and Pople
argued that the low-lying excited states of the $\pi$-conjugated
systems can be described by an effective $\pi$-electron model Hamiltonian,
now known as the PPP Hamiltonian \cite{pople1953electron,pariser1953semi},
which can be written in the second-quantized form as
\begin{equation}
H=\sum_{i,j,\sigma}t_{ij}(c_{i\sigma}^{\dagger}c_{j\sigma}+c_{j\sigma}^{\dagger}c_{i\sigma})+U\sum_{i}n_{i\uparrow}n_{i\downarrow}+\sum_{i<j}V_{ij}(n_{i}-1)(n_{j}-1),\label{eq:ppp}
\end{equation}

where $c_{i\sigma}^{\dagger}(c_{i\sigma})$ is the creation (annihilation)
operator which creates (annihilates) a $\pi$-electron with spin $\sigma$
localized on the $i^{th}$ carbon atom, while the number of $\pi$
electrons with spin $\sigma$ is indicated by $n_{i\sigma}=c_{i\sigma}^{\dagger}c_{i\sigma}$,
and $n_{i}=\sum_{\sigma}n_{i\sigma}$ denotes the total number of
$\pi$ electrons on the $i^{th}$ carbon atom. Each carbon atom contributes
a single $\pi$ orbital to the basis set, which, for the systems lying
in the $xy$ plane, is nothing but the $p_{z}$ orbital of the atom.
In the first term of Eq. \ref{eq:ppp}, $t_{ij}$ represents the one
electron hopping matrix element, while $U$ and $V_{ij}$ in the second
and third terms denote the onsite, and long-range Coulomb interactions,
respectively. Similar to our previous calculations on $\pi$-conjugated
materials, we only considered nearest-neighbor hopping matrix elements
to be non-zero \cite{shukla2002correlated,shukla2004theory,chakraborty2013pariser,chakraborty2014theory,bhattacharyya2020pariser}
whose values were determined using the formula proposed by Ramasesha
and coworkers\cite{rama-thio}
\begin{equation}
t_{ij}=-2.4+3.2(R_{i,j}-1.397),\label{eq:tij}
\end{equation}
 where $R_{i,j}$ is the distance between the $i^{th}$ and $j^{th}$
carbon atoms in $\text{Å}$ units. Note that, unlike the first-principles
approach, in the PPP model hydrogen atoms are not considered because
they participate only in the $\sigma$ bonds.

Ohno relationship is used to parameterize the Coulomb interactions
\cite{ohno1964some} 
\begin{equation}
V_{ij}=U/\kappa_{i,j}(1+0.6117R_{i,j}^{2})^{1/2},
\end{equation}

where $\kappa_{i,j}$ denotes dielectric constant of the system, using
which we can include the screening effects, and $R_{i,j}$ is the
same as in Eq. \ref{eq:tij}. The positions of various carbon atoms
required to compute $R_{i,j}$, were obtained by performing geometry
optimization using a first-principles DFT approach discussed in the
previous section. In this work, PPP calculations were performed using
two sets of Coulomb parameters: (a) the standard parameters {[}$U=11.13\text{ eV},\kappa_{i,j}=1.0]$,
and (b) the screened parameters {[}$U=8.0$ eV, $\kappa_{i,j}$= 2.0
($i\neq j$) and $\kappa_{i,i}=1.0${]} \cite{chandross1997coulomb}.
The use of screened Coulomb parameters yields superior agreement with
the experiments performed in the condensed phase because they simulate
the screening effects due to the host, i.e., the solvent or solid-state
matrix.

\subsubsection{CI Calculations: Computational Steps}

A calculation is started by performing a restricted Hartree-Fock (RHF)
calculation for the closed-shell singlet ground states of the PAH
molecules considered here, by using a computer program implementing
the PPP model, developed in our group \cite{sony2010general}. The
molecular orbitals (MOs) obtained from the RHF calculations are next
used to transform the PPP Hamiltonian from the site basis, to the
molecular orbital (MO) basis, for performing electron-correlated calculations
using the configuration interaction (CI) approach. Next, symmetry
and spin-adapted singles-doubles CI (SDCI) calculations are performed
using the computer program MELD \cite{mcmurchie1990meld}, to compute
the ground and the electric-dipole allowed excited state energies
and many-electron wave functions. Subsequently, the transition dipole
moments connecting the ground to the excited states are computed and
used to evaluate the optical absorption spectrum of the molecule employing
the formula
\begin{equation}
\sigma(\omega)=4\pi\alpha\sum_{n}\frac{\omega_{ng}|\left\langle n|\hat{e}.\boldsymbol{r}|g\right\rangle |^{2}\gamma^{2}}{(\omega_{ng}-\omega)^{2}+\gamma^{2}},\label{eq:opt-spec}
\end{equation}

where $|g\rangle$ ($|n\rangle$) denotes the ground (excited) state
CI wave function, $\hbar\omega_{ng}=E_{n}-E_{g}$ is the energy difference
between the two states, $\left\langle n|\hat{e}.\boldsymbol{r}|g\right\rangle $
is the corresponding transition dipole matrix element for an incident
photon of energy $\hbar\omega$ polarized along the $\hat{e}$ direction,
$\alpha$ denotes the fine-structure constant, and $\gamma$ is the
assumed uniform line width for all the excited states. Next we examine
the excited states contributing to the prominent peaks in the calculated
spectrum, and include the dominant configuration state functions contributing
to them in our list of reference configurations for performing the
multi-reference singles-doubles CI (MRSDCI) calculations in the next
step. Subsequently, the absorption spectrum at the MRSDCI level is
calculated and compared to the one computed in the previous CI calculation.
This procedure is iterated until the absorption spectrum converges
within an acceptable tolerance \cite{shukla2002correlated,shukla2004theory,chakraborty2013pariser,chakraborty2014theory,bhattacharyya2020pariser}. 

\section{Results and Discussion}

The point group symmetries of dibenzo{[}bc,ef{]}coronene, dinaphtho{[}8,1,2abc;2´,1´,8´klm{]}coronene,
and dinaphtho{[}8,1,2abc;2´,1´,8´jkl{]}coronene respectively are $C_{2v}$,
$C_{2v}$, and $C_{2h}$. Therefore, in order to represent the PAHs
considered in this work succinctly, we have adopted a notation of
the form GQD-$N$-$PG$, where $N$ denotes the total number of carbon
atoms in the PAH, and $PG$ is its point group which can be $C_{2v}$
or $C_{2h}$ in the present case. Therefore, henceforth, dibenzo{[}bc,ef{]}coronene,
dinaphtho{[}8,1,2abc;2´,1´,8´klm{]}coronene, and dinaphtho{[}8,1,2abc;2´,1´,8´jkl{]}coronene
will be denoted as GQD-30-$C_{2v}$, GQD-36-$C_{2v}$, and \textsubscript{}
GQD-36-$C_{2h}$, respectively.

\label{sec:Results-and-Discussion}

\subsection{Optimized Structures of the GQDs}

\begin{figure}[H]
\begin{centering}
\includegraphics[scale=0.5]{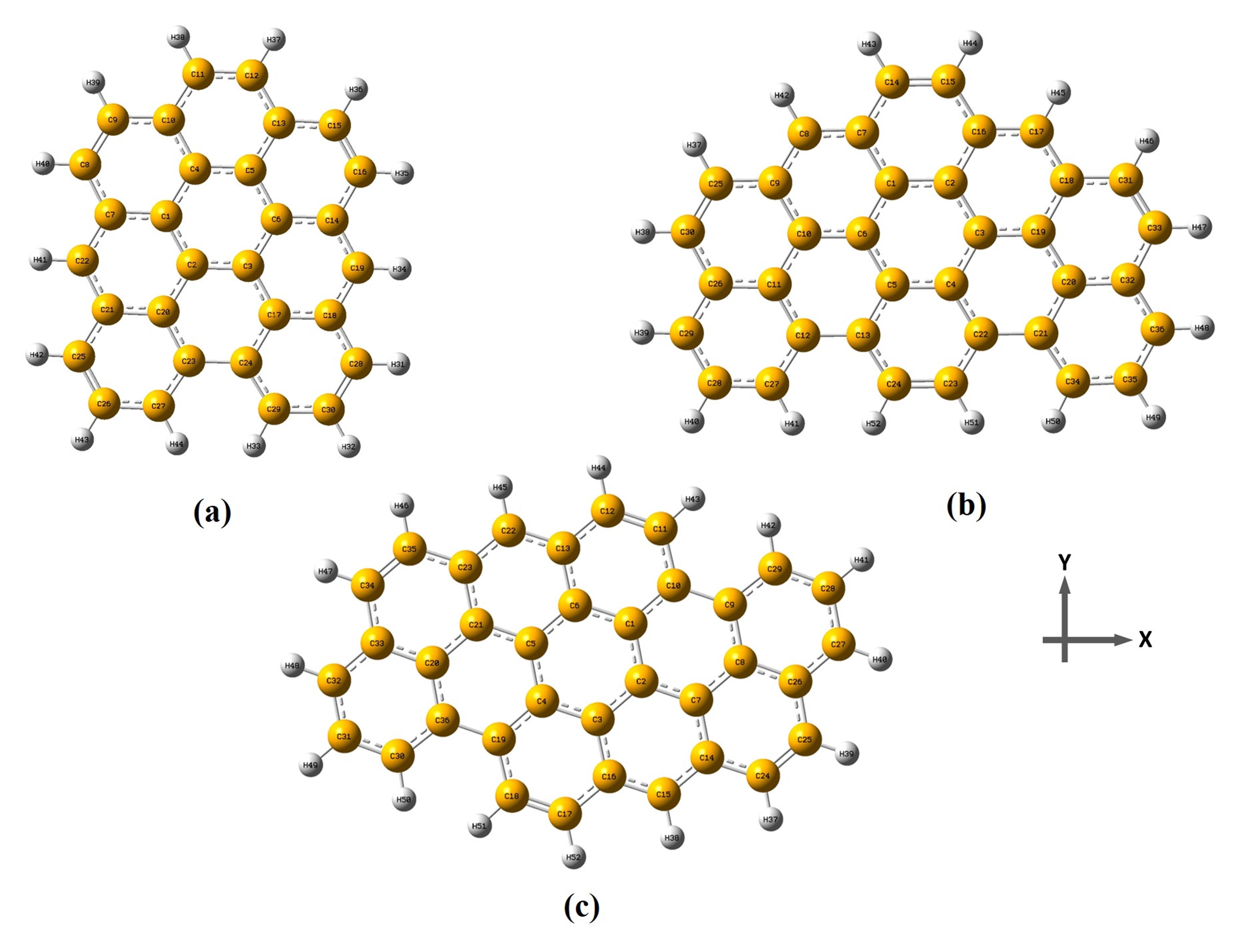} 
\par\end{centering}
\caption{Optimized geometries of the three GQDs: (a) GQD-30-$C_{2v}$, (b)
GQD-36-$C_{2v}$, and (c) GQD-36-$C_{2h}$. In the figures, yellow/grey
spheres denote C/H atoms. \protect\label{fig:opt-geom}}
\end{figure}

Optimized geometries of the three GQDs considered in this work are
shown in Fig. \ref{fig:opt-geom}. All the final structures are strictly
planar, with the bond lengths and bond angles showing variations close
to the ideal values of 1.4 Å and 120°, respectively. For the first
molecule (GQD-30-$C_{2v}$), the minimum and maximum bond lengths
are 1.39 $\text{Å}$ (C11-C12) and 1.44 $\text{Å}$ (C17-C18), while
the minimum and maximum bond angles are 119.80° (C2-C3-C6) and 120.60°
(C28-C30-C29), respectively. The parentheses following the bond lengths
and angles list the carbon atoms involved in those, according to the
atom numbering scheme in Fig. \ref{fig:opt-geom}(a). For GQD-36-$C_{2v}$
(Fig. \ref{fig:opt-geom}(b)), the minimum and maximum bond lengths
and angles are 1.38 $\text{Å}$ (C23-C24), 1.43 Å (C1-C7), and 119.91°
(C18-C19-C20, C9-C10-C11), 120.30° (C32-C36-C35, C28-C29-C26), respectively.
Finally, for GQD-36-$C_{2h}$ (Fig. \ref{fig:opt-geom}(c)), the corresponding
minimum and maximum values are 1.39 Å (C27-C28) , 1.43 Å (C8-C9),
and 119.64° (C5-C13-C22), 120.36° (C31-C32-C33). For all the three
GQDs, all C-H bond lengths at the edges were very close to 1.09 Å
. Furthermore, the geometries of the three GQDs even after the optimization
process retain the point group symmetries.

\subsection{Electronic Structure}

\begin{figure}[H]
\begin{centering}
\includegraphics[scale=0.6]{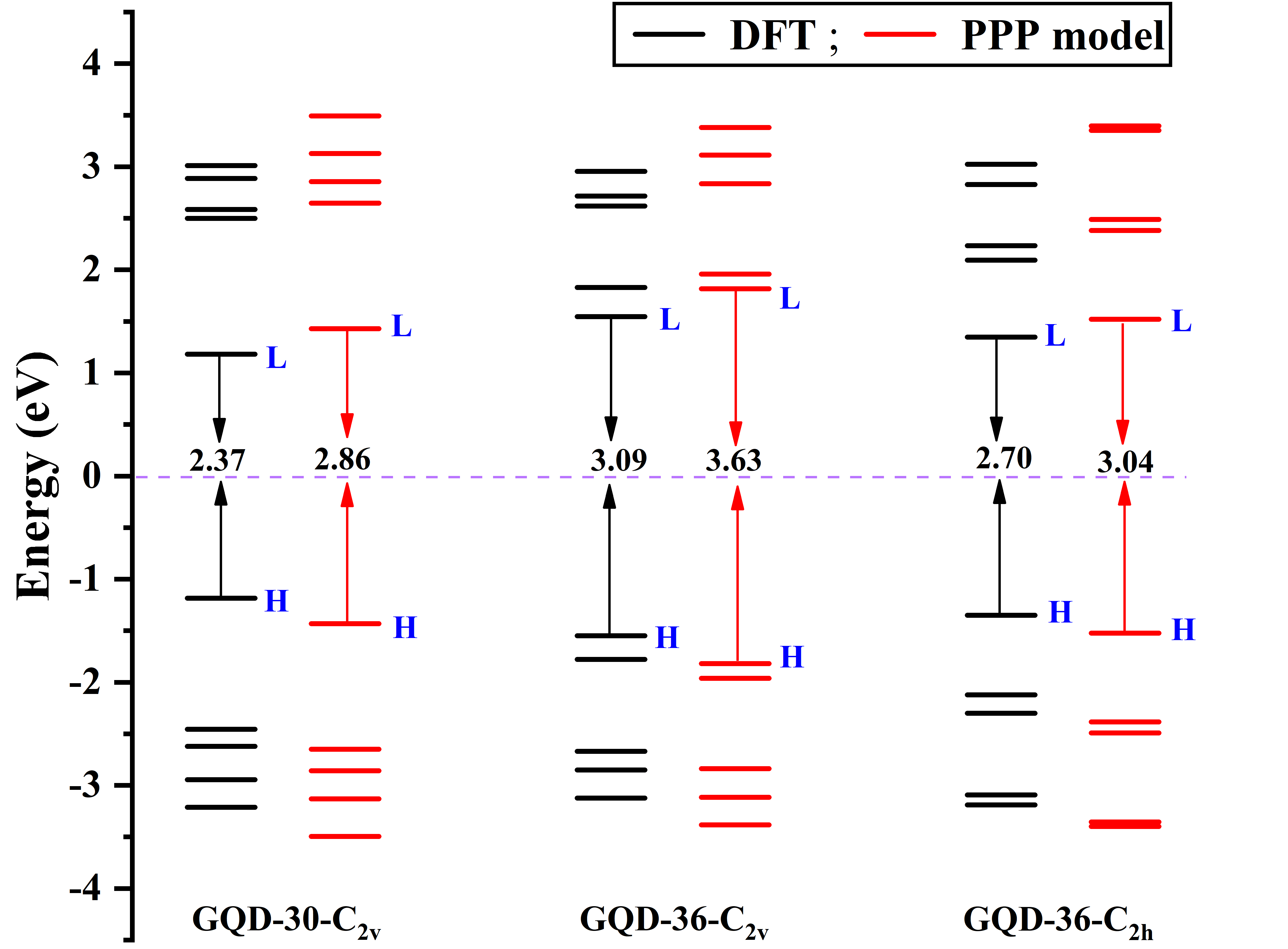}
\par\end{centering}
\caption{Energy levels corresponding to HOMO$-$4 to LUMO$+$4 along with the
HOMO$-$LUMO energy gaps of all the three GQDs computed using the
DFT, and PPP-HF calculations employing the screened parameters.\protect\label{fig:e-level}
We shifted the Fermi level to 0 eV for both sets of orbitals for the
ease of comparison.}
\end{figure}

First we briefly discuss the electronic structure of the three GQDs
at the one-electron level computed using the DFT as well as PPP-Hartree-Fock
(PPP-HF) levels corresponding to the optimized geometries. In Fig.
\ref{fig:e-level} we plot the electronic energy levels for the orbitals
close to the Fermi level obtained using the DFT as well as the PPP-HF
(screened parameters) approaches, while Fig. \ref{fig:DOS} contains
the total and atom-projected density of states (DOS) computed at the
DFT level. In Fig. \ref{fig:e-level} we note that for each molecule,
the pattern of the energy levels obtained from the DFT and PPP-HF
approaches are quite similar. However, due to the well-known tendency
of the DFT to underestimate the energy gaps, the HOMO-LUMO gaps obtained
using the DFT are significantly smaller than those computed using
the PPP-HF method. 

On examining the atom-projected DOS presented in Fig. \ref{fig:DOS},
we note that for energies near and lower than those of HOMO, carbon
atoms contribute dominantly to it, which also holds true for a significant
energy range near and above LUMO. An examination of the molecular
orbitals near the Fermi level reveals that they have strictly $\pi$
and $\pi*$ character, which is also confirmed by the orbital-projected
DOS plots presented in the Fig. S1 of Supporting Information. Therefore,
the optical absorption spectrum of these GQDs is dominated by $\pi\rightarrow\pi^{*}$
excitations, in full agreement with the assumptions involved in the
PPP model.
\begin{flushleft}
\begin{figure}[H]
\begin{raggedright}
\includegraphics[width=16.3cm]{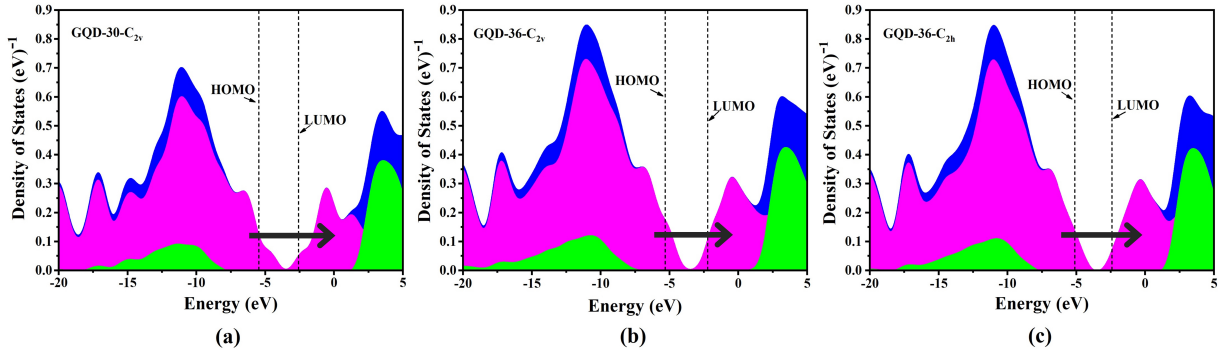}
\par\end{raggedright}
\caption{The total and atom-projected density of states (DOS) of the three
hydrogen passivated graphene quantum dots (a) GQD-30-C\protect\textsubscript{2v},
(b) GQD-36-C\protect\textsubscript{2v}, (c) GQD-36-C\protect\textsubscript{2h}.
The regions in blue color represent the total density of states (TDOS),
while the magenta and green colors denote the contributions of C and
H atoms, respectively, . HOMO and LUMO are represented with vertical
dashed lines. Big horizontal arrow between the occupied and unoccupied
states represents the $\pi\rightarrow\pi^{*}$transitions responsible
for the dominant optical absorption of the three GQDs.\protect\label{fig:DOS}}
\end{figure}
\par\end{flushleft}

\subsection{Size of the PPP-CI Calculations}

All the GQDs considered in this work are charge neutral, as a result
they correspond to half-filled systems with each carbon atom contributing
a single $\pi$ electron. Therefore, with the total number of electrons
in the GQDs being 30 or 36, all of them have a closed-shell singlet
ground state corresponding to the symmetric irreducible representation
(irrep) of their point group. As a result, the ground-state symmetry
of GQD-30-$C_{2v}$ and GQD-36-$C_{2v}$ is $^{1}A_{1}$, while that
of GQD-36-$C_{2h}$ is $^{1}A_{g}$. Within the PPP approach, one
can only consider optically excited states corresponding to the transitions
caused by photons polarized in the plane of the molecule. Therefore,
according to the dipole selection rules of the point groups $C_{2v}$
and $C_{2h}$, the possible irreps of the one-photon excited states
are either $^{1}A_{1}$ or $^{1}B_{2}$ for GQD-30-$C_{2v}$, GQD-36-$C_{2v}$,
and $^{1}B_{u}$ for GQD-36-$C_{2h}$.

For the calculations of the optical gaps and absorption spectra of
various GQDs, a coefficient cutoff value of 0.05 was used in the MRSDCI
calculations. This means that all those configurations in the MRSDCI
expansion of the targeted state (ground or the excited state) with
the coefficients larger in magnitude than 0.05 are included in the
list of the reference configurations in the next iteration of the
MRSDCI calculation, and the procedure is repeated until the desired
quantities converge. For the calculation of the spin gap, i.e., the
energy gap between the singlet ground state and lowest-energy triplet
state of a GQD, only the two states in question were targeted in the
MRSDCI calculations and the calculations were continued until the
spin gap converged. As a result we included configurations with a
cutoff value 0.03 for the coefficients, leading to significantly larger
CI expansions as compared to those needed for the calculation of the
absorption spectra. To illustrate the large-scale nature of our CI
calculations, the total number of spin-adapted configurations ($N_{total}$),
i.e., the dimension of the CI matrix for the largest calculations
of a given symmetry manifold are presented in Table \ref{tab:ci-ntotal}.
From the Table it is obvious that the size of the MRSDCI expansion
ranged from close to one million to almost five million, suggesting
that our calculations are well-converged.

\begin{table}[H]
\caption{Total number of symmetry and spin-adapted configurations ($N_{total}$)
used in the largest PPP-MRSDCI calculations for different symmetries,
employing the standard (std) and screened (scr) Coulomb parameters.\protect\label{tab:ci-ntotal}}

\centering{}%
\begin{tabular}{cccc}
\toprule 
GQD & Irrep & $N_{total}$(std) & $N_{total}$(scr)\tabularnewline
\midrule
\midrule 
GQD-30-$C_{2v}$ & $^{1}A_{1}$ & $\text{1283990}${\small{} } & 1590944\tabularnewline
 & $^{1}B_{2}$ & 839374 & 971738\tabularnewline
 & $^{3}B_{2}$ & $\text{2221359}$  & $\text{2641743}$\tabularnewline
GQD-36-$C_{2v}$ & $^{1}A_{1}$ & $\text{1978364}$ & $\text{3075530}$\tabularnewline
 & $^{1}B_{2}$ & $\text{1427001}$ & $\text{1940654}$\tabularnewline
 & $^{3}B_{2}$ & $\text{3855805}$ & $\text{4078607}$\tabularnewline
GQD-36-$C_{2h}$ & $^{1}A_{g}$ & $\text{3056665}$ & $\text{3304115}$\tabularnewline
 & $^{1}B_{u}$ & $\text{2314242}$ & $\text{1729422}$\tabularnewline
 & $^{3}B_{u}$ & $\text{4601107}$ & $\text{4820343}$\tabularnewline
\bottomrule
\end{tabular}
\end{table}

\subsection{Optical Gaps}

We will discuss the computed absorption spectra of the three GQDs
in detail in the next section, however, first we discuss their calculated
and measured optical gaps, corresponding to the first peaks. In Table
\ref{tab:optical-gaps}, we present the results of our calculations
for the optical gaps of the three GQDs, along with the corresponding
experimental results \cite{fetzer2000large,clar1977correlations,BRYSON20111980}.
We note that both for the TDDFT as well as PPP-CI calculations, the
optical gap corresponds to a many-electron state dominated by HOMO-LUMO
singly-excited configuration denoted as $|H\rightarrow L\rangle$.
As far as quantitative values are concerned, the gaps computed using
the PPP-CI approach coupled with the screened parameters are, on the
average, in the best agreement with the solution-based experiments
of Clar and Schmidt \cite{clar1977correlations} and Fetzer \cite{fetzer2000large}.
However, for the case of GQD-36-$C_{2v}$, thin-film based measurements
of Bryson \emph{et al}. \cite{BRYSON20111980} are lower than all
the calculated as well as reported values of Fetzer \cite{fetzer2000large}.
The possible reason behind this may be the coupling between different
molecules (chromophores) in the thin-film phase, leading to a red
shift in the first absorption peak as compared to the solution-based
results. The gaps computed using the standard parameters in the PPP-CI
method overestimate the experimental values for all the cases, while
those computed using the first-principles TDDFT method underestimate
it for GQD-30-$C_{2v}$. However, the TDDFT results for GQD-36-$C_{2v}$
and GQD-36-$C_{2h}$ are in very good agreement with the experiment. 

\begin{table}[H]
\caption{Optical gaps of the three GQDs calculated using the TDDFT method along
with PPP-CI approach employing the standard (std) and screened (scr)
parameters. The experimental results are also presented for comparison.\protect\label{tab:optical-gaps}}

\centering{}%
\begin{tabular}{>{\centering}p{6cm}c>{\centering}p{1.5cm}>{\centering}p{1.5cm}c}
\toprule 
GQD & Optical gap (eV) & \multicolumn{2}{c}{Optical gap (eV)} & Optical gap (eV)\tabularnewline
 & TDDFT & \multicolumn{2}{c}{PPP-CI} & Experiment\tabularnewline
\midrule
\midrule 
 &  & scr & std & \tabularnewline
\midrule 
GQD-30-$C_{2v}$ & 2.22 & 2.36 & 2.76 & 2.46 \cite{clar1977correlations}\tabularnewline
 &  &  &  & \tabularnewline
GQD-36-$C_{2v}$ & 2.79 & 2.91 & 3.22 & 2.77 \cite{BRYSON20111980}, 2.88 \cite{fetzer2000large}\tabularnewline
 &  &  &  & \tabularnewline
GQD-36-$C_{2h}$ & 2.51 & 2.53 & 2.97 & 2.50 \cite{fetzer2000large}\tabularnewline
\bottomrule
\end{tabular}
\end{table}

\subsection{Optical Absorption Spectra }

In this section, we present and analyze the computed linear optical
absorption spectra, obtained using both the first-principles TDDFT
approach followed by those computed using the PPP model and the MRSDCI
approach (PPP-CI approach, in short) for the three PAH molecules considered
in this work. For our calculations, the optimized geometries presented
in Fig. \ref{fig:opt-geom} were utilized. For the chosen orientation
of the Cartesian axes, in $C_{2v}$ GQDs, an optical transition from
the ground state to a $^{1}B_{2}$ excited states will be through
an $x$-polarized photon, while a transition to a state of $^{1}A_{1}$
symmetry will involve a $y$-polarized photon. In the $C_{2h}$ GQD,
an optical transition from the ground state will be to a state of
$^{1}B_{u}$ symmetry, through a photon of mixed $x,y$ polarization.

On comparing the PPP-CI spectra of each molecule computed using the
standard and the screened parameters, we see the following qualitative
similarities: (a) generally speaking, the peaks and features in the
spectrum calculated using the screened parameters are at lower excitation
energies as compared to the one computed using the standard parameters,
(b) the first peaks of the spectra are of weak to moderate intensities
corresponding to the optical gap, with the excited states dominated
by the $|H\rightarrow L\rangle$ configuration, and (c) for this state,
the absorbed photon is $x$-polarized for the GQDs of the $C_{2v}$
symmetry, and has mixed polarization for the $C_{2h}$ symmetric GQD.

Having already discussed the locations of the first peaks (optical
gap) in the absorption spectra in the previous section, here we focus
on the higher energy peaks, and, in particular, the maximum intensity
(MI) peaks, whose excitation energies are presented in Table \ref{tab:gqd-mi}.
\begin{center}
\begin{table}[H]
\caption{\protect\label{tab:gqd-mi}The positions of the most intense (MI)
peaks in the optical absorption spectra of the GQDs considered in
this work, computed using different approaches, compared to the corresponding
experimental values. The peak number and polarization direction of
the MI peaks are indicated in the parentheses. }

\centering{}%
\begin{tabular}{>{\centering}p{3cm}cccc}
\toprule 
GQD & \multicolumn{4}{c}{Most intense peak (eV)}\tabularnewline
\midrule 
 & TDDFT & PPP-CI (scr) & PPP-CI (std) & Experiment (eV)\tabularnewline
\midrule 
GQD-30-$C_{2v}$ & 3.83 (III\textsubscript{$y$}) & 3.41 (II\textsubscript{$y$}) & 4.29 (II\textsubscript{$y$}) & 3.74 \cite{clar1977correlations}\tabularnewline
GQD-36-$C_{2v}$ & 3.34 (II\textsubscript{$xy$}) & 3.17 (II\textsubscript{$xy$}) & 3.94 (II\textsubscript{$xy$}) & 3.29 \cite{BRYSON20111980}, 3.39 \cite{fetzer2000large}\tabularnewline
GQD-36-$C_{2h}$ & 3.40 (II\textsubscript{$xy$}) & 3.22 (II\textsubscript{$xy$}) & 4.05 (II\textsubscript{$xy$}) & 3.41 \cite{fetzer2000large}\tabularnewline
\bottomrule
\end{tabular}
\end{table}
\par\end{center}

The detailed information regarding the excitations and the wave functions
contributing to the peaks of the optical absorption spectra of all
the three molecules are presented in tables S1-S9 of the Supporting
Information (SI).

\subsubsection{GQD-30-$C_{2v}$ }

\subsubsection*{TDDFT Spectrum}

\begin{figure}[H]
\begin{centering}
\includegraphics[width=14cm,height=5cm]{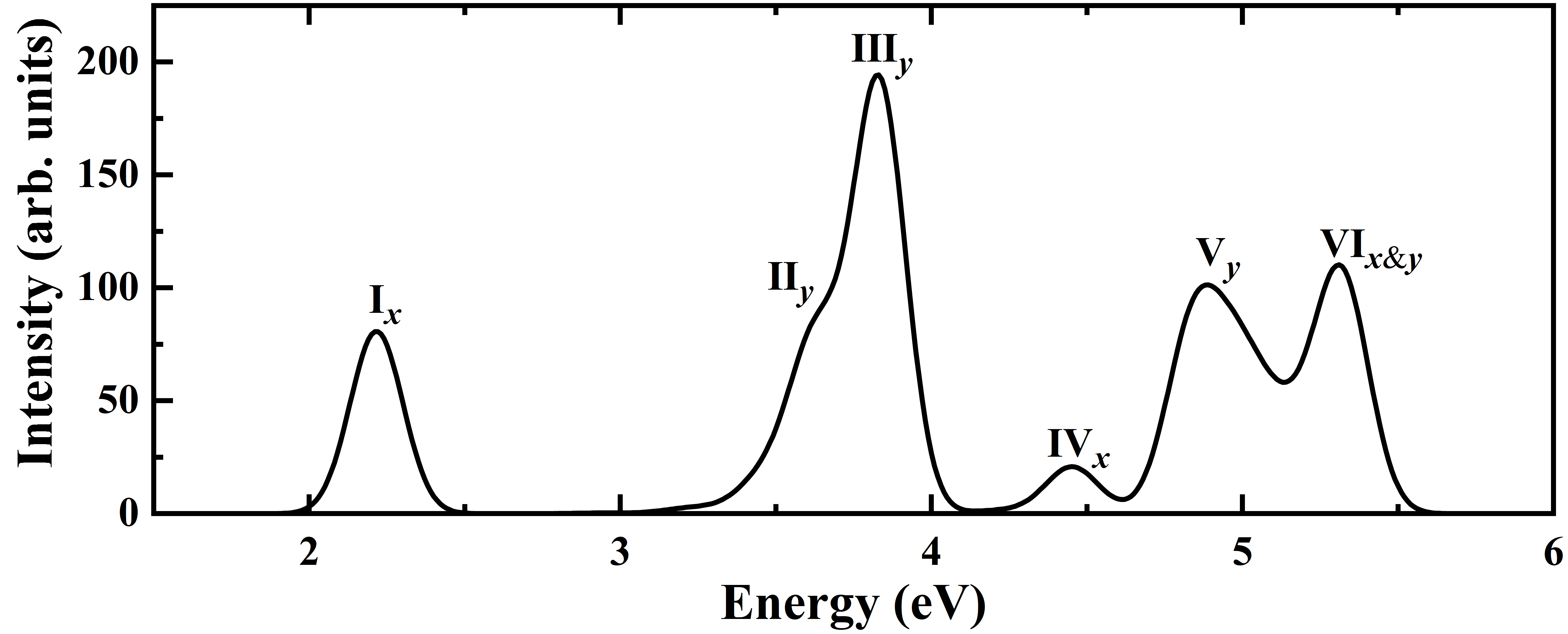}
\par\end{centering}
\caption{UV-Vis absorption spectrum of GQD-30-$C_{2v}$, i.e., dibenzo{[}bc,ef{]}coronene
(C\protect\protect\textsubscript{30}H\protect\protect\textsubscript{14}),
computed using the first-principles TDDFT method using a uniform line
width of 0.1 eV. The subscript associated with each peak label indicates
the direction of polarization of the absorbed photon. \protect\label{fig:gqd-30-tddft}}
\end{figure}

In Fig. \ref{fig:gqd-30-tddft} we present the linear optical absorption
spectrum of GQD-30-$C_{2v}$ computed using the TDDFT approach coupled
with the B3LYP functional. The spectrum has six features, of which
the first peak located at 2.22 eV is due to the absorption of a photon
with polarization along the $x$ direction, to a state whose wave
function is dominated by the $|H\rightarrow L\rangle$ configuration.
As far as the intensity profile of the absorption spectrum is concerned,
the first peak is not the most intense, rather it is the $y$-polarized
third peak located at 3.83 eV which is the most intense one. The experimentally
measured location of the most intense peak at 3.74 eV \cite{clar1977correlations}
is in a very good agreement with this value. The dominant configuration
contributing to the excited state giving rise to this peak (III) is
$|H\rightarrow L+3\rangle$, while those contributing to the shoulder
II and the weak peak IV are $|H-3\rightarrow L\rangle$, and $|H-1\rightarrow L+1\rangle$,
respectively. Two more significant peaks V and VI are there in the
higher energy region at 4.87 eV and 5.32 eV respectively. Dominant
configurations giving rise to y polarized peak V are $|H-2\rightarrow L+2\rangle$,
$|H-2\rightarrow L+5\rangle$, while peak VI shows both x and y polarizations
and has contributions from two nearly degenerate states of $^{1}B_{2}$
and $^{1}A_{1}$ symmetries whose wave functions are dominated by
single excitations $|H-2\rightarrow L+5\rangle$, and $|H-4\rightarrow L+3\rangle$,
respectively. The detailed information about the excited states contributing
to the peaks of the spectrum are presented in Table S1 of Supporting
Information. 

\subsubsection*{PPP-CI Spectra}

\begin{figure}[H]
\begin{centering}
\includegraphics[width=14cm,height=9cm]{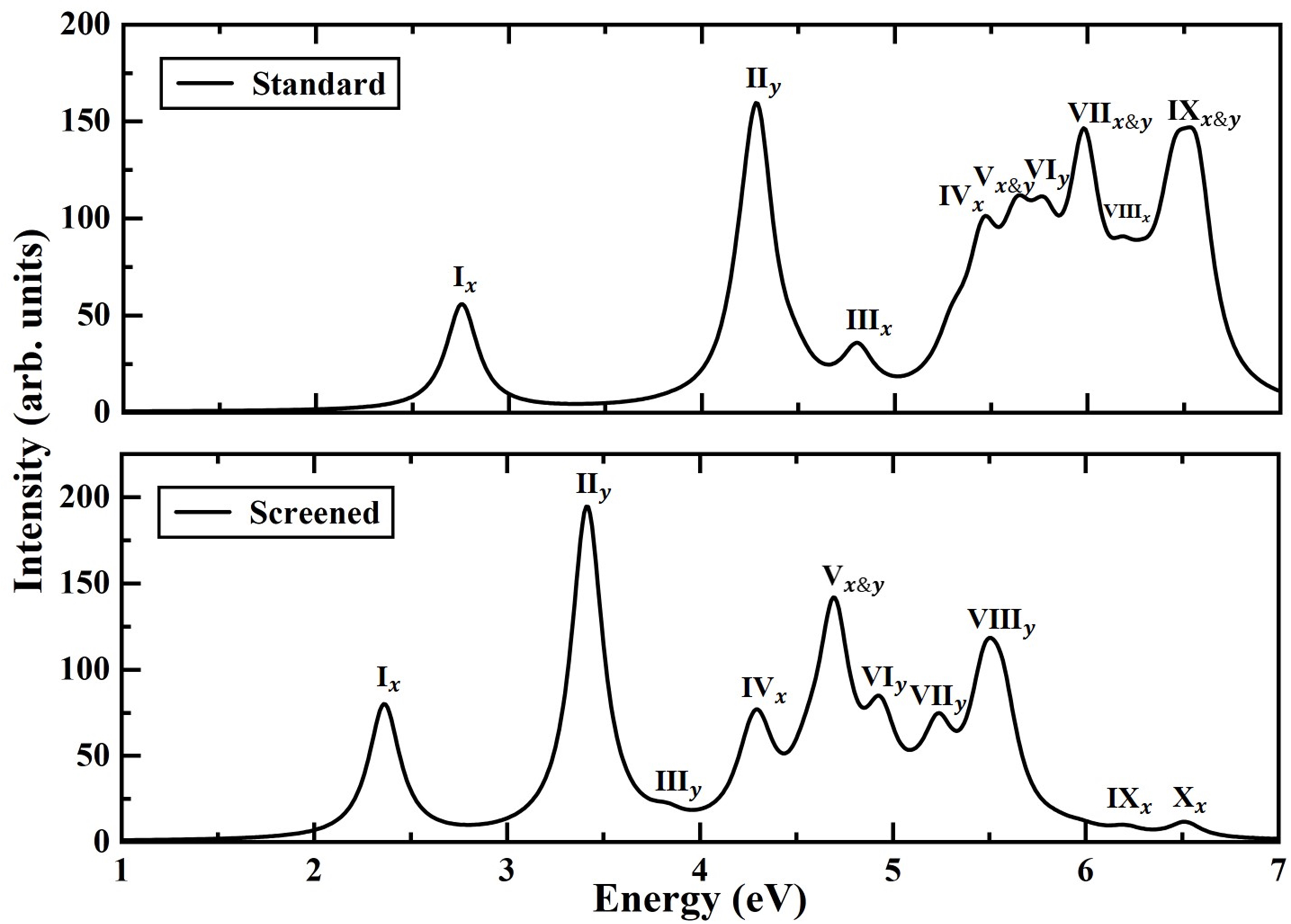} 
\par\end{centering}
\caption{\protect\label{fig:ppp-spectra-gqd-30}Linear optical absorption spectra
of GQD-30-$C_{2v}$, i.e., dibenzo{[}bc,ef{]}coronene (C\protect\protect\textsubscript{30}H\protect\protect\textsubscript{14})
computed using the PPP-CI methodology. The calculations were performed
using the standard parameters (upper panel) and the screened parameters
(lower panel) with a uniform line width of 0.1 eV. Polarization directions
of the transitions are specified by the subscripts of the peak labels.}
\end{figure}

We present PPP-CI level calculated linear optical absorption spectra
of GQD-30-$C_{2v}$ employing both the standard and the screened Coulomb
parameters in Fig. \ref{fig:ppp-spectra-gqd-30}. For both the calculations,
the second peak (II) corresponding to a $y$-polarized transition
is the MI peak, with the excited states dominated by the singly excited
configurations $|H-1\rightarrow L\rangle+c.c.$, where $c.c.$ denotes
the charge-conjugated configuration $|H\rightarrow L+1\rangle$. As
far as quantitative comparison among the approaches on location of
the MI peak is concerned (see Table \ref{tab:gqd-mi}), the PPP-CI
value 3.41 eV computed using the screened parameters is about 0.4
eV lower than the location of peak III at 3.83 eV in TDDFT spectrum.
However, if we consider the MI peak to be a broad band starting at
peak II in the TDDFT spectrum located at 3.63 eV, the difference reduces
to about 0.2 eV. The PPP-CI location of the MI peak based on the standard
parameter calculations at 4.29 eV, is about 0.5 eV higher than the
TDDFT value. Compared to the experimental location of the MI peak
at 3.74 eV \cite{clar1977correlations}, our screened parameter value
is about 0.3 eV lower while the standard parameter value is about
0.6 eV higher.

After the MI peak, the next significant high intensity peak in the
screened parameter spectrum occurs at 4.70 eV (peak V) due to two
almost degenerate excited states with $x$ and $y$ polarizations.
The wave function of the $y$-polarized state is dominated by single
excitations $|H-6\rightarrow L\rangle+c.c.$ while the $x$-polarized
state is mainly composed of the single excitation $\arrowvert H-2\rightarrow L+2\rangle$
with small contribution from the triple excitation $\arrowvert H\rightarrow L;H\rightarrow L;H-2\rightarrow L+2\rangle$
indicating some influence of electron-correlation effects. The $y$-polarized
peak VIII located at 5.48 eV also carries significant oscillator strength
and is largely due to the single excitations $\arrowvert H-1\rightarrow L+5\rangle+c.c.$.

In the standard parameter spectrum, beyond the MI peak, there is a
absorption band of width $\approx$ 1 eV starting near 5.5 eV and
ending close to 6.5 eV with several closely-spaced high intensity
peaks. Of these, peaks VII and IX carry maximum intensities, and exhibit
both $x$ and $y$ polarizations due to nearly degenerate states of
\textsuperscript{1}A\textsubscript{1 }and \textsuperscript{1}B\textsubscript{2}
symmetries. Peak VII is dominated by states with single excitations
$|H-6\rightarrow L\rangle+c.c.$, $\arrowvert H-2\rightarrow L+2\rangle$,
and $\arrowvert H-3\rightarrow L+3\rangle$, while single excitations
$|H-9\rightarrow L\rangle+c.c.$, $\arrowvert H-4\rightarrow L+4\rangle$,
$\arrowvert H-1\rightarrow L+5\rangle+c.c.$, and the triple excitation
$\arrowvert H\rightarrow L;H-4\rightarrow L+1;H-1\rightarrow L+4\rangle$
give rise to peak IX. Detailed information about all the peaks in
the PPP-CI spectra with standard and screened parameters are given
in Table S4 and S5 of Supporting Information.

If we compare the two sets of spectra up to the excitation energy
range 4.5 eV, we note that both the TDDFT and PPP-CI (screened) spectra
have four features each, while PPP-CI (standard) has just two features.
Furthermore, the polarization directions of those four peaks in TDDFT
and PPP-CI (screened) spectra match perfectly with each other. There
is also a decent quantitative agreement in the peak locations predicted
by the two spectra. For example, peak IV of TDDFT is located at 4.45
eV while the same peak of PPP-CI (screened) spectrum is at 4.30 eV.
Also, the TDDFT peaks V (4.87 eV) and VI (5.32 eV) are comparable
to the PPP-CI (screened) peaks VI and VII at 4.93 eV and 5.23 eV,
respectively. Therefore, we conclude that the first-principles TDDFT
spectrum is in good qualitative and quantitative agreement with that
computed by the PPP-CI (screened) approach.

\subsubsection{GQD-36-$C_{2v}$ }

\subsubsection*{TDDFT Spectrum}

\begin{figure}[H]
\begin{centering}
\includegraphics[width=14cm,height=5cm]{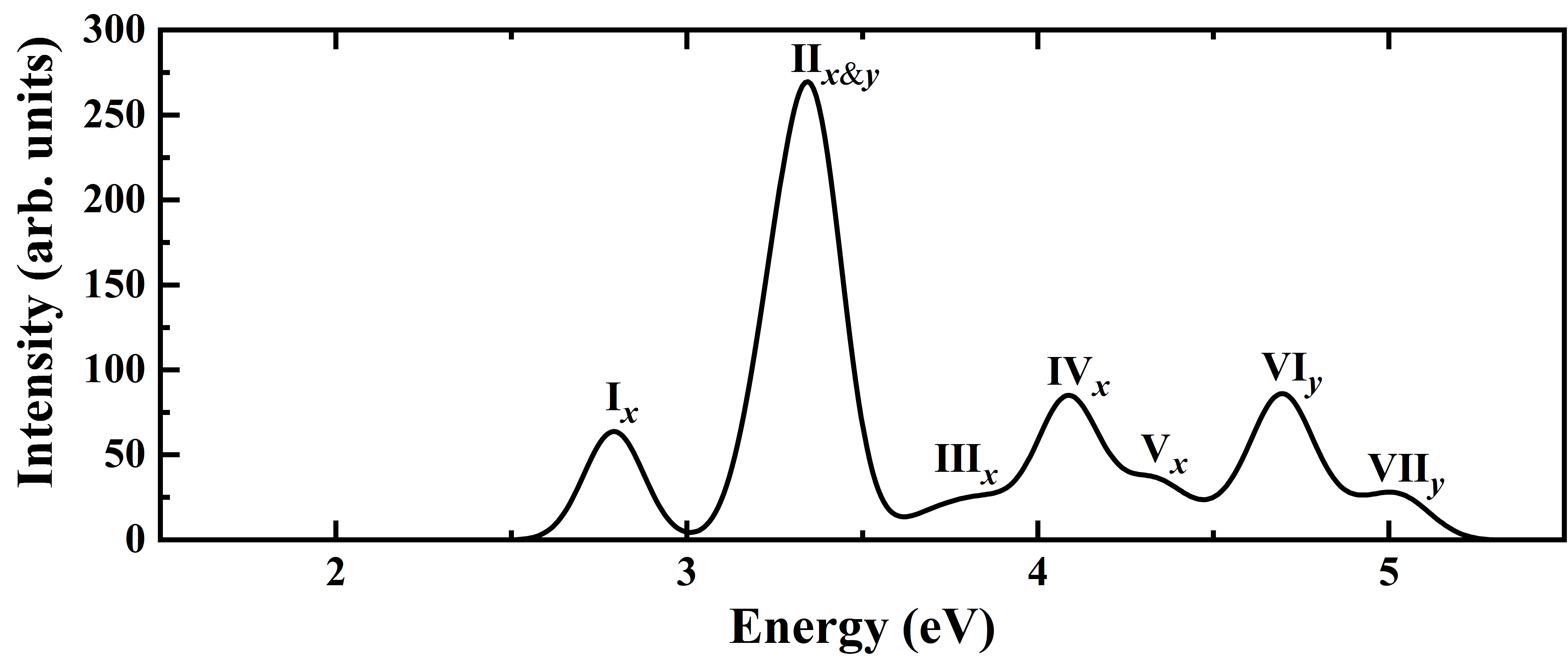}
\par\end{centering}
\caption{UV-Vis absorption spectrum of GQD-36-$C_{2v}$, i.e., dinaphtho{[}8,1,2abc;2´,1´,8´klm{]}coronene
(C\protect\protect\textsubscript{36}H\protect\protect\textsubscript{16}),
computed using the first-principles TDDFT method using a uniform line
width of 0.1 eV. The subscript associated with each peak label indicates
the direction of polarization of the absorbed photon.\protect\label{fig:36-c2v-tddft}}
\end{figure}

We present the linear optical absorption spectrum of GQD-36-$C_{2v}$,
computed using the TDDFT approach, in Fig. \ref{fig:36-c2v-tddft}.
We have identified seven features in the spectrum with the maximum
excitation energy $\approx5$ eV. The first peak corresponding to
the optical gap is due to an excited state whose many-body wave function
is dominated by the $|H\rightarrow L\rangle$ excitation, and similar
to the case of GQD-30-$C_{2v}$, it is also reached by the absorption
of an $x$-polarized photon. The most intense peak of the computed
spectrum is peak II located at 3.34 eV shows both $x$ and $y$ polarizations
due to two closely-spaced excited states of symmetries $^{1}B_{2}$
and $^{1}A_{1}$ whose wave functions are dominated by single excitations
$|H-1\rightarrow L+1\rangle$, and $|H\rightarrow L+1\rangle$, respectively.
To the oscillator strength of this peak, the $x$-polarized $^{1}B_{2}$
contributes significantly more as compared to the $y$-polarized $^{1}A_{1}$
state. In the region beyond the MI peak, there are five weaker peaks
of which the $x$-polarized peak IV near 4.07 eV and $y$-polarized
peak VI at 4.70 carry significant intensity, and are dominated by
the excitations $|H-2\rightarrow L+1\rangle$ and $|H-2\rightarrow L+1\rangle$,
respectively. The rest of the peaks (III, V, VII) are quite feeble
and of comparable intensity, occur at 3.86 eV, 4.33 eV and 5.04 eV,
respectively. The detailed information about the excited states contributing
to the peaks of the spectrum of this GQD can be found in Table S2
of Supporting Information. 

\subsubsection*{PPP-CI Spectra}

\begin{figure}[H]
\begin{centering}
\includegraphics[width=14cm,height=9cm]{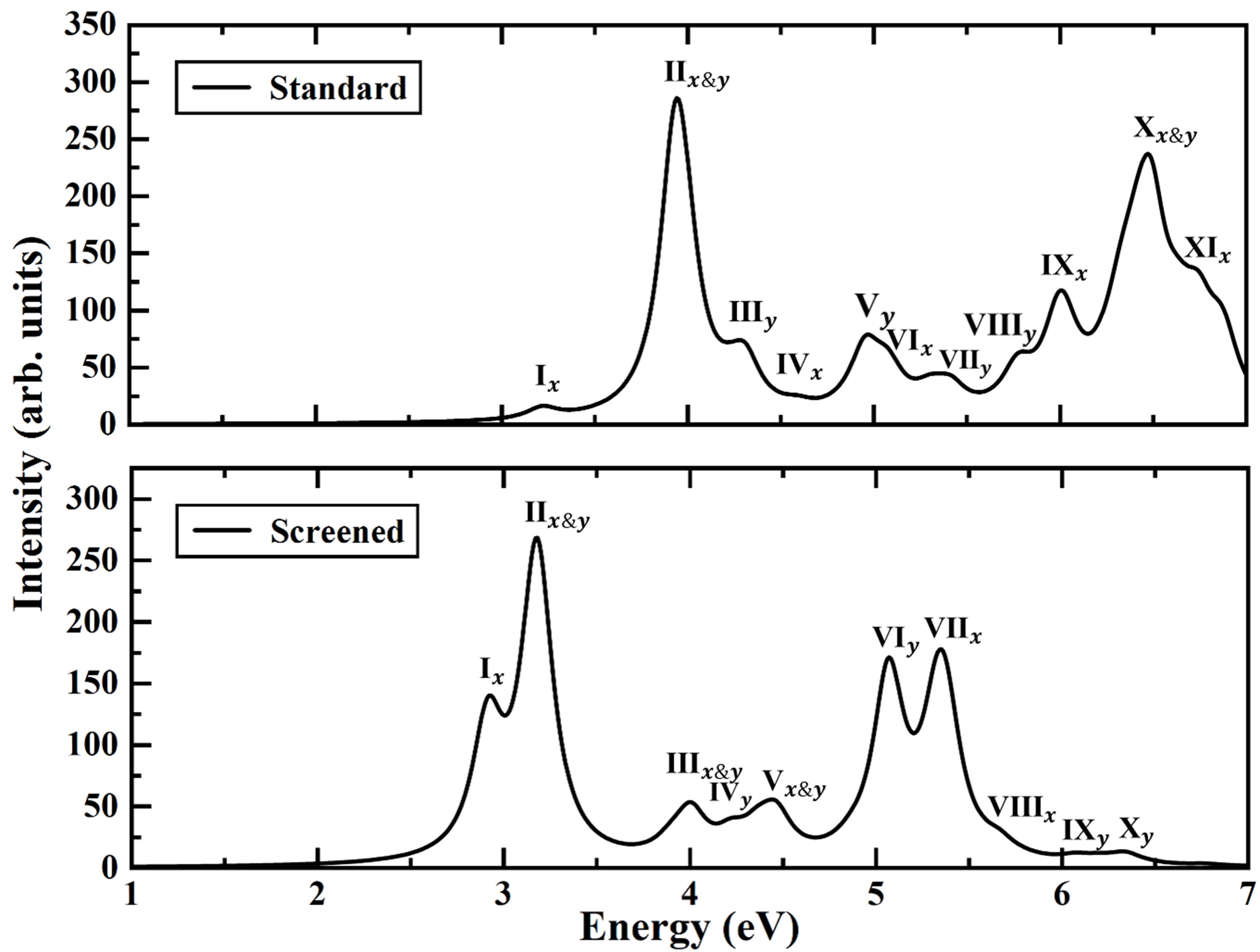} 
\par\end{centering}
\caption{Linear optical absorption spectra of GQD-36-$C_{2v}$, i.e., dinaphtho{[}8,1,2abc;2´,1´,8´klm{]}coronene
(C\protect\protect\textsubscript{36}H\protect\protect\textsubscript{16})
computed using the PPP-CI methodology. The calculations were performed
using the standard parameters (upper panel) and the screened parameters
(lower panel) with a uniform line width of 0.1 eV. Polarization directions
of the transitions are specified by the subscripts of the peak labels.
\protect\label{fig:ppp-spectra-gqd-36-c2v}}
\end{figure}

We present the linear optical absorption spectra of GQD-36-$C_{2v}$
computed using the PPP-CI approach, employing both sets of Coulomb
parameters, in Fig. \ref{fig:ppp-spectra-gqd-36-c2v}. The first peak
in the spectrum computed using the standard parameters is of very
low intensity while the corresponding peak in the screened parameter
spectrum is of moderate intensity. 

In both the spectra, the second peak is the MI peak, which derives
large oscillator strengths from two nearly degenerate excited states
of $^{1}B_{2}$ and $^{1}A_{1}$ symmetries, whose wave functions
are dominated by single excitations $\arrowvert H-1\rightarrow L+1\rangle$,
and $\arrowvert H\rightarrow L+1\rangle+c.c.$, respectively. The
PPP-CI (screened) location of the MI peak at 3.17 eV is slightly lower
than the corresponding TDDFT value of 3.34 eV, while the standard
parameter location of the MI peak at 3.94 eV is significantly higher
than it (see Table \ref{tab:gqd-mi}).

In the computed spectrum of PPP-CI (screened) there are two more peaks
VI (5.07 eV) and VII (5.33 eV) of significant intensities, that are
$y$- and $x$-polarized, with their many-body wave functions dominated
by the single excitations $\arrowvert H-3\rightarrow L+2\rangle+c.c.$
and $\arrowvert H-4\rightarrow L+3\rangle+c.c.$, respectively. Additionally,
there are several other weaker features in the computed spectrum at
energies higher than 5.5 eV.

In the PPP-CI (standard) spectrum, the MI peak is followed by a number
of weak-intensity peaks. The next intense peak (X) is of mixed polarization
and occurs due to three closely-spaced states at 6.32 eV ($^{1}B_{2}$),
6.41 eV ($^{1}B_{2}$), and 6.49 eV ($^{1}A_{1}$), dominated by excitations
$\arrowvert H-3\rightarrow L+3\rangle$, $\arrowvert H-2\rightarrow L+5\rangle+c.c.$,
and $\arrowvert H-2\rightarrow L+3\rangle+c.c.$. The detailed information
about the excited states contributing to various peaks can be found
in Tables S6 (standard) and S7 (screened) of the Supporting Information.

We note that the TDDFT spectrum of this molecule has seven features
in all extending up to slightly beyond 5 eV (see Fig. \ref{fig:36-c2v-tddft}).
In that energy region PPP-CI spectra both with screened and standard
parameters have six features. The polarization characteristics of
the first two peaks are in agreement in all the three spectra, and
the intensity profiles also match in this energy region. However,
beyond the MI peak and below 5 eV, the peaks are of low to moderate
intensities.

\subsection*{Comparison with the experimental data}

For GQD-36-$C_{2v}$ (dinaphtho{[}8,1,2abc;2´,1´,8´klm{]}coronene),
we found two previously measured experimental spectra, one from solution-based
experiment reported by Fetzer \cite{fetzer2000large}, and other one
from the thin-film based experiment of Bryson \emph{et} \textit{al}.
\cite{BRYSON20111980}. We have re-plotted the experimental spectra
in energy unit (eV) which was originally reported in the unit of wavelength
(nm) and compared it with our theoretically computed spectra {[}see
Fig. \ref{fig:Comparison-of-theory-expt-36-c2v}{]}. Both the calculated
absorption spectra obtained using the TDDFT approach, and the PPP-CI
method using the screened parameters, show good agreement with both
the experimental spectra. The first peak of our TDDFT spectrum is
at 2.79 eV which agrees well with the peak at 2.77 eV of the experimental
spectrum of Bryson et al. {[}Fig. 7(b){]} \cite{BRYSON20111980},
while the first peak of PPP-CI spectrum (with screened parameters)
at 2.91 eV is very close to the peak at 2.88 eV of the experimental
spectra of Fetzer {[}see Fig. 7(a){]} \cite{fetzer2000large}. Thus,
these experimental and theoretical peaks are within the energy range
2.77 -- 2.91 eV, i.e., a spread of just $\approx0.14$ eV, which
signifies good agreement. As far as the location of the MI peaks of
the calculated spectrum are concerned, the TDDFT value (3.34 eV) is
again in good agreement with the experimental peaks near 3.29 eV and
3.39 eV reported by Bryson et al. \cite{BRYSON20111980}, and Fetzer
\cite{fetzer2000large}, respectively. TDDFT peak at 4.70 eV is in
excellent agreement with experimental peaks near 4.7 eV \cite{fetzer2000large,BRYSON20111980}.
The MI peak of PPP-CI spectrum at 3.17 eV is also quite close to the
experimental peaks at 3.29 eV (deviated by $\sim$ 0.1 eV) by Bryson
et al. and 3.39 eV (deviated by $\sim$ 0.2 eV) by Fetzer. We note
that in the measurement of Bryson \emph{et al}. \cite{BRYSON20111980}
a continuous increase in the intensity is observed beyond 4 eV, which
is absent both in our theoretical results as well in the experiment
of Fetzer.\cite{fetzer2000large} In the region below that energy,
we note generally a good qualitative and quantitative agreement between
both the experimental and theoretical results.
\begin{center}
\begin{figure}[H]
\begin{centering}
. \includegraphics[scale=0.7]{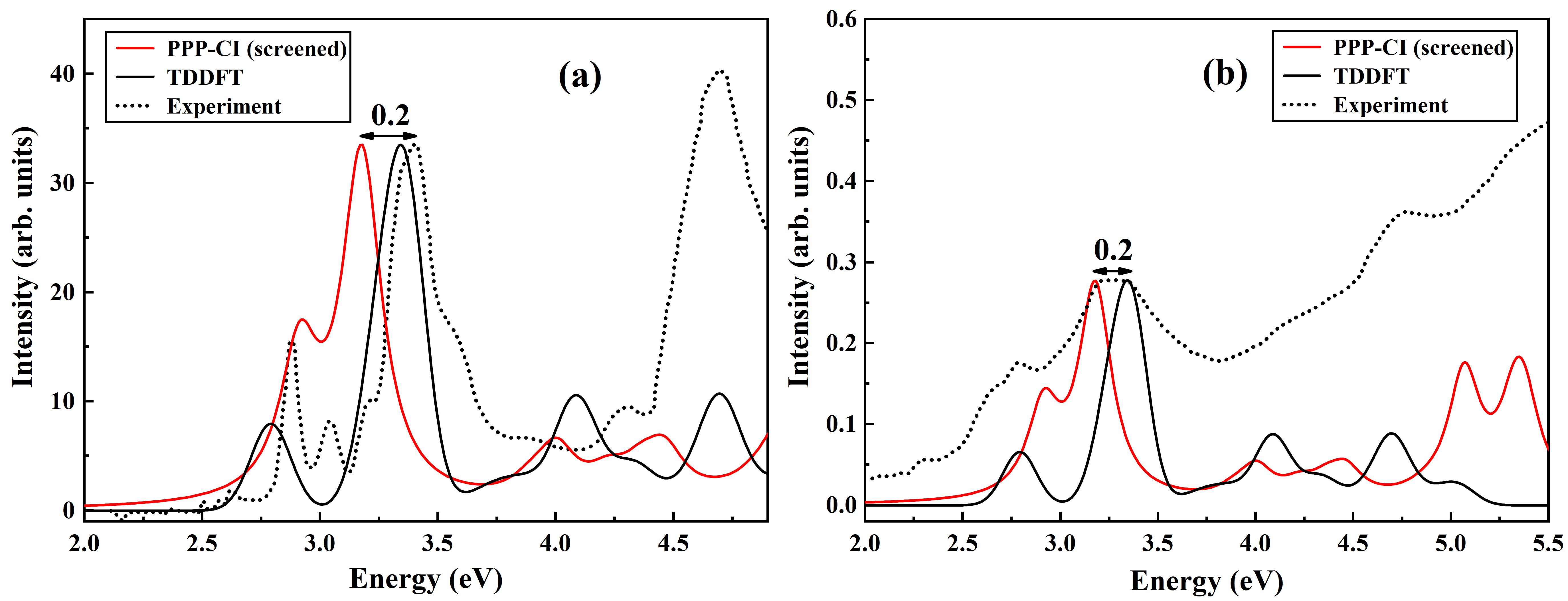}
\par\end{centering}
\caption{The comparison of the linear optical absorption spectra of GQD-36-$C_{2v}$
(dinaphtho{[}8,1,2abc;2´,1´,8´klm{]}coronene) computed using the TDDFT
(solid black line) and the screened parameter PPP-CI (solid red line)
approaches, with the experimental spectra (dotted line) (a) reproduced
with permission from ref. \cite{fetzer2000large} Copyright 2001,
(John Wiley and Sons) and (b) reproduced with permission from ref.
\cite{BRYSON20111980} Copyright 2011, (Elsevier). The calculated
spectra have been scaled so that their MI peaks have the same intensity
as a nearby experimental peak. We have converted the original experimental
spectra reported in the wavelength (nm) units to energy (eV) units
for direct comparison with our calculations. \protect\label{fig:Comparison-of-theory-expt-36-c2v}}
\end{figure}
 
\par\end{center}

In the PPP-CI spectrum, beyond the MI peak there are several features
of which the peaks near 4.0 and 4.4 eV are close to experimental features
around 3.9 and 4.3 eV, respectively, in the spectrum of Fetzer \cite{fetzer2000large}.
Similarly, if we compare with the experimental spectrum of Bryson
et al.\cite{BRYSON20111980}, we can see from Fig. \ref{fig:Comparison-of-theory-expt-36-c2v}(b)
that a feature near 4.8 eV is close to a PPP-CI peak near 5.1 eV.
Furthermore, Bryson et al.\cite{BRYSON20111980} in the discussion
of their spectrum have mentioned a shoulder near 4.3 eV (287 nm),
which is close to a PPP-CI feature at 4.4 eV and a TDDFT peak at 4.1
eV. The experiment of Fetzer \cite{fetzer2000large} has the most
intense peak at 4.7 eV which disagrees with both the TDDFT and PPP-CI
spectra that report peaks of lesser intensities in that region.

\subsubsection{GQD-36-$C_{2h}$ }

\subsubsection*{TDDFT Spectrum}

\begin{figure}[H]
\begin{centering}
\includegraphics[width=14cm,height=5cm]{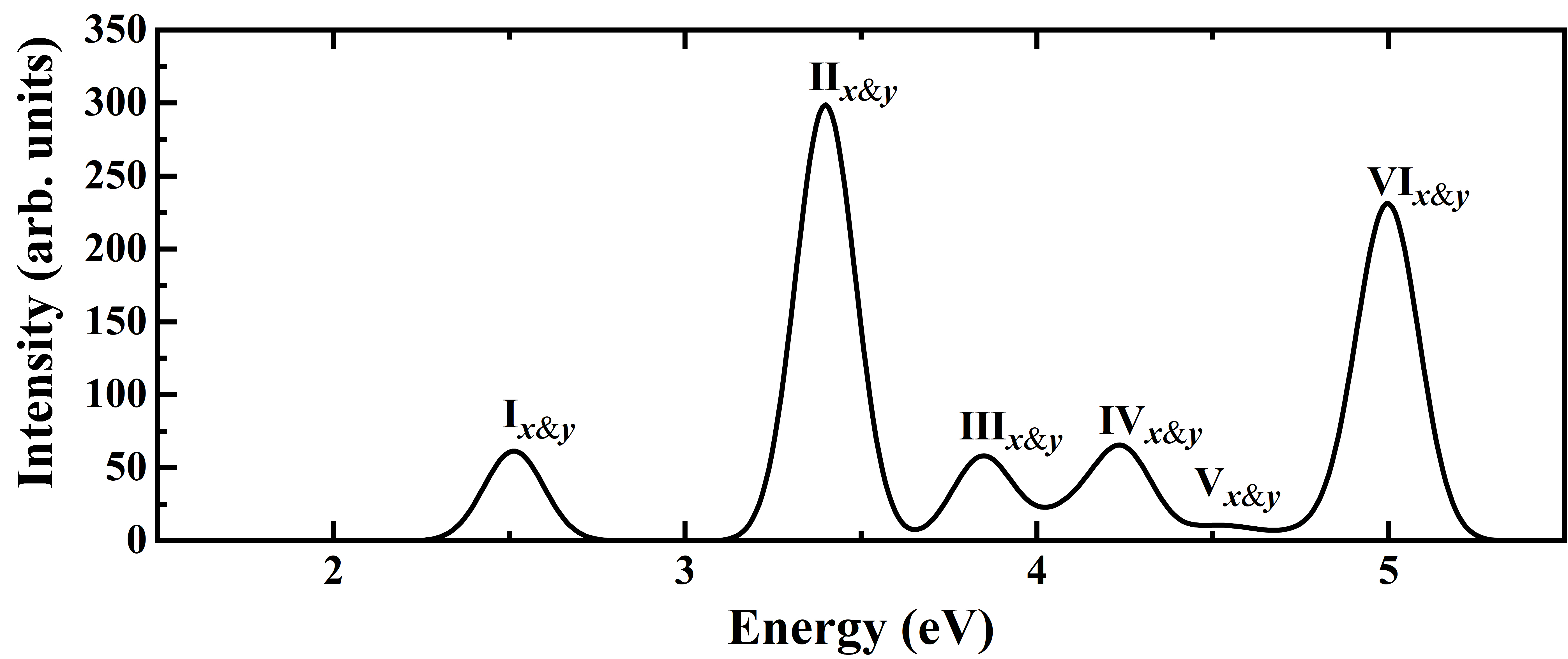}
\par\end{centering}
\caption{UV-Vis absorption spectrum of GQD-36-$C_{2h}$, i.e., dinaphtho{[}8,1,2abc;2´,1´,8´jkl{]}coronene
(C\protect\protect\textsubscript{36}H\protect\protect\textsubscript{16}),
computed using the first-principles TDDFT method using a uniform line
width of 0.1 eV. The subscript associated with each peak label indicates
the direction of polarization of the absorbed photon.\protect\label{fig:36-c2h-tddft}}
\end{figure}

Fig. \ref{fig:36-c2h-tddft} presents the linear optical absorption
spectrum of GQD-36-$C_{2h}$ computed using the TDDFT approach and
the B3LYP functional. The spectrum displays five well-defined peaks,
all of which have mixed polarization characteristics in agreement
with the electric-dipole selection rules of the $C_{2h}$ point group.
The first peak located at 2.51 eV is due to a state whose wave function
is dominated by the $|H\rightarrow L\rangle$ configuration, while
the second peak at 3.40 eV is the MI peak whose wave function is mainly
composed of the single excitation $|H-1\rightarrow L\rangle$. The
third and the fourth peaks of the spectrum located at 3.84 and 4.24
eV are as intense as the first peak, and are due to excited states
dominated by single excitations $\arrowvert H-1\rightarrow L+1\rangle$,
and $\arrowvert H-2\rightarrow L+2\rangle$, respectively. After the
MI peak one high intensity peak (VI) appears near 5 eV due to two
closely spaced excitations at 4.97 eV and 5.01 eV dominated by configurations
$\arrowvert H-3\rightarrow L+2\rangle$, and $\arrowvert H-2\rightarrow L+6\rangle$,
respectively.

More details about the excited states contributing to the spectrum
can be found in Table S3 of Supporting Information. 

\subsubsection*{PPP-CI Spectra}

\begin{figure}[H]
\begin{centering}
\includegraphics[width=14cm,height=9cm]{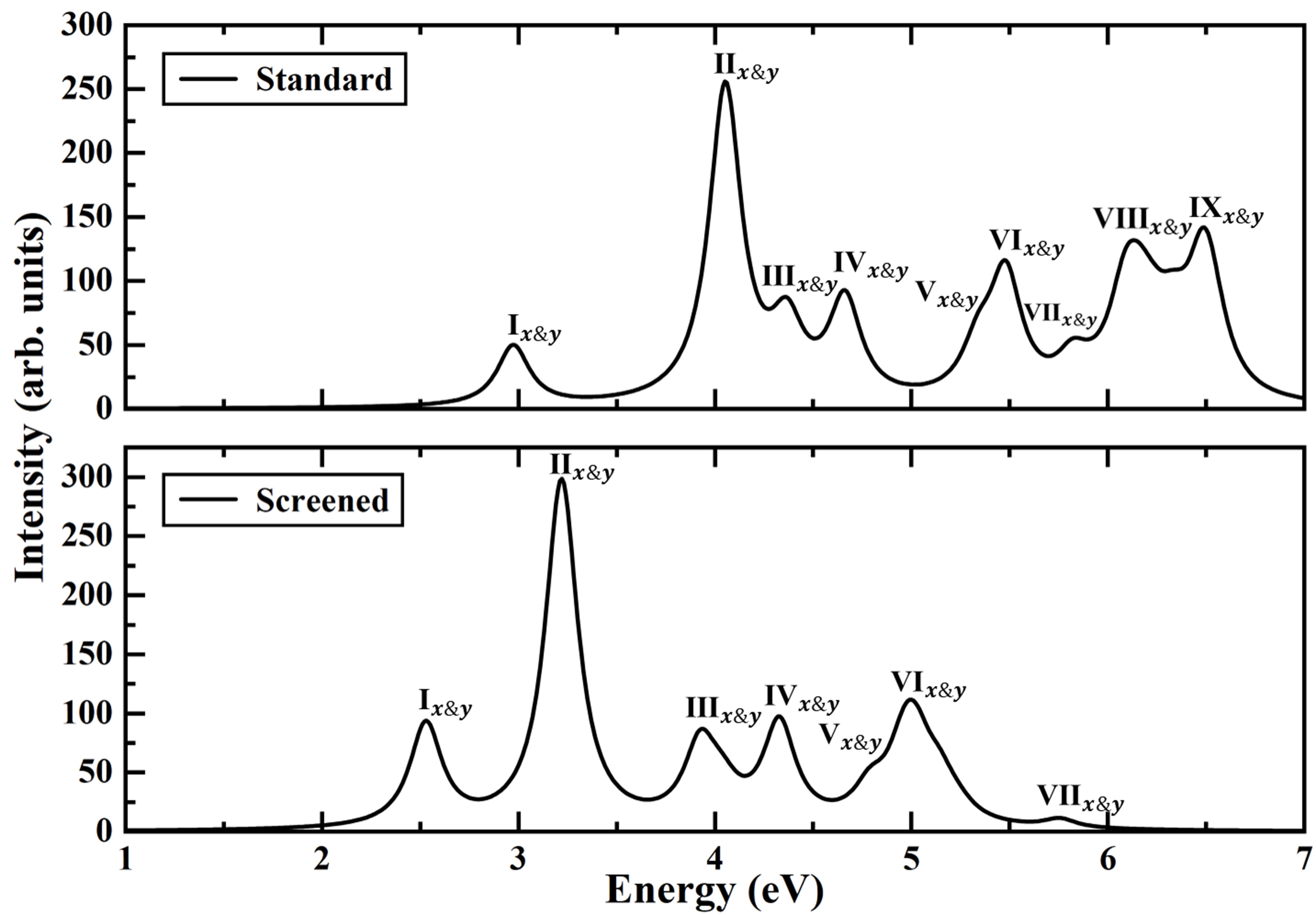} 
\par\end{centering}
\caption{Linear optical absorption spectra of GQD-36-$C_{2h}$, i.e., dinaphtho{[}8,1,2abc;2´,1´,8´jkl{]}coronene
(C\protect\protect\textsubscript{36}H\protect\protect\textsubscript{16})
computed using the PPP-CI methodology. The calculations were performed
using the standard parameters (upper panel) and the screened parameters
(lower panel) with a uniform line width of 0.1 eV. Polarization directions
of the transitions are specified by the subscripts of the peak labels.\protect\label{fig:ppp-spectra-gqd-36-c2h}}
\end{figure}

The optical absorption spectra of GQD-36-$C_{2h}$ computed using
the PPP-CI approach, employing both the standard and screened Coulomb
parameters are presented in Fig. \ref{fig:ppp-spectra-gqd-36-c2h}.
The first peak in the spectrum computed using the standard parameters
has lower relative intensity as compared the corresponding peak in
the screened parameter spectrum. 

The MI peak is the second peak (II) in both the spectra, and the state
giving rise to it is dominated by $\arrowvert H\rightarrow L+1\rangle+c.c.$
singly-excited configurations. As far the location of the MI peak
is concerned, the PPP-CI (screened) value of 3.22 eV is slightly smaller
than the TDDFT value 3.40 eV, while the PPP-CI (standard) location
at 4.05 eV is significantly larger than it (see Table \ref{tab:gqd-mi})

In the standard parameter spectrum, beyond the MI peak there are several
moderate-intensity peaks all the way up to 6.5 eV. Out of these, two
peaks (III and IV) are below 5 eV, while the remaining five features
are above that energy. The wave functions of the states leading to
peak III (4.37 eV) and peak IV (4.66 eV) exhibit significant configuration
mixing. Peak III derives strong contributions from single excitations
$\arrowvert H-2\rightarrow L+2\rangle$ and $\arrowvert H-1\rightarrow L+1\rangle$,
while configurations $\arrowvert H-1\rightarrow L+1\rangle$ and $\arrowvert H-5\rightarrow L\rangle+c.c.$
dominate the state giving rise to peak IV. With the increasing energy
of the peaks, the transitions occur to orbitals farther away from
the Fermi level. For example, the peaks V (5.34 eV) and IX (6.47 eV)
are dominated by excitations $\arrowvert H\rightarrow L+8\rangle+c.c.$,
and $\arrowvert H-4\rightarrow L+4\rangle$, respectively. 

In the screened parameter spectrum, beyond the MI peak, there are
three more moderate intensity peaks (III, IV, and VI) in the region
up to 5.0 eV, while the last peak (VII) at 5.75 eV is a very faint
one. The excited states contributing to peaks III (3.93 eV), IV (4.33
eV), and VI (5.03 eV) are dominated by the excitations $\arrowvert H-1\rightarrow L+1\rangle$,
$\arrowvert H-2\rightarrow L+2\rangle$, and $\arrowvert H-5\rightarrow L+1\rangle+c.c.$,
respectively. More information about the excited states contributing
to the PPP-CI spectra can be found in Tables S8 and S9 of the Supporting
Information for the standard and screened parameter calculations,
respectively.

The TDDFT spectrum of this molecule has six peaks in all, the last
of which is located near 5 eV (see Fig. \ref{fig:36-c2h-tddft}).
In that energy range, broadly speaking PPP-CI (screened) spectrum
also has six peaks (peak VI is at 5.03 eV), while the standard parameter
spectrum has four peaks. Furthermore, peak III of TDDFT spectrum located
at 3.84 eV compares very well with the peak III of the screened parameter
spectrum at 3.93 eV. Peak VI near 5 eV of both the TDDFT and PPP-CI
(screened) spectra are in excellent agreement with each other. Thus,
we conclude that the spectra computed by TDDFT and the PPP-CI (screened)
approaches are in excellent qualitative and quantitative agreement
with each other for GQD-36-$C_{2h}$. 

\subsection*{Comparison with the experimental data}

In case of GQD-36-$C_{2h}$ also the optical spectra computed using
PPP-CI (screened parameters) and the TDDFT approaches show good agreement
with the experimental spectra reported by Fetzer \cite{fetzer2000large}.
Here also we have re-plotted the experimental spectrum in energy unit
(eV) which was originally reported in the wavelength unit (nm) and
compared with our theoretically computed spectra in Fig. \ref{fig:Comparison-of-theory-expt-36-c2h}.
In the figure we can see that the first peak in the lower energy region
of the spectrum at 2.53 eV predicted by the PPP-CI approach is very
close to the peak at 2.50 eV in the experimental spectrum. Furthermore,
the MI peak of the PPP-CI spectrum at 3.22 eV is just 0.2 eV separated
from the experimental MI peak at 3.41 eV indicating good agreement.
We also note that in the experimental spectrum \cite{fetzer2000large},
peaks at 3.85 eV, 4.30 eV, and 4.70 are very close to the locations
of our calculated PPP-CI peaks at 3.93 eV, 4.33 eV, and the shoulder
at 4.79 eV, respectively.
\begin{center}
\begin{figure}[H]
\begin{centering}
\includegraphics[scale=0.4]{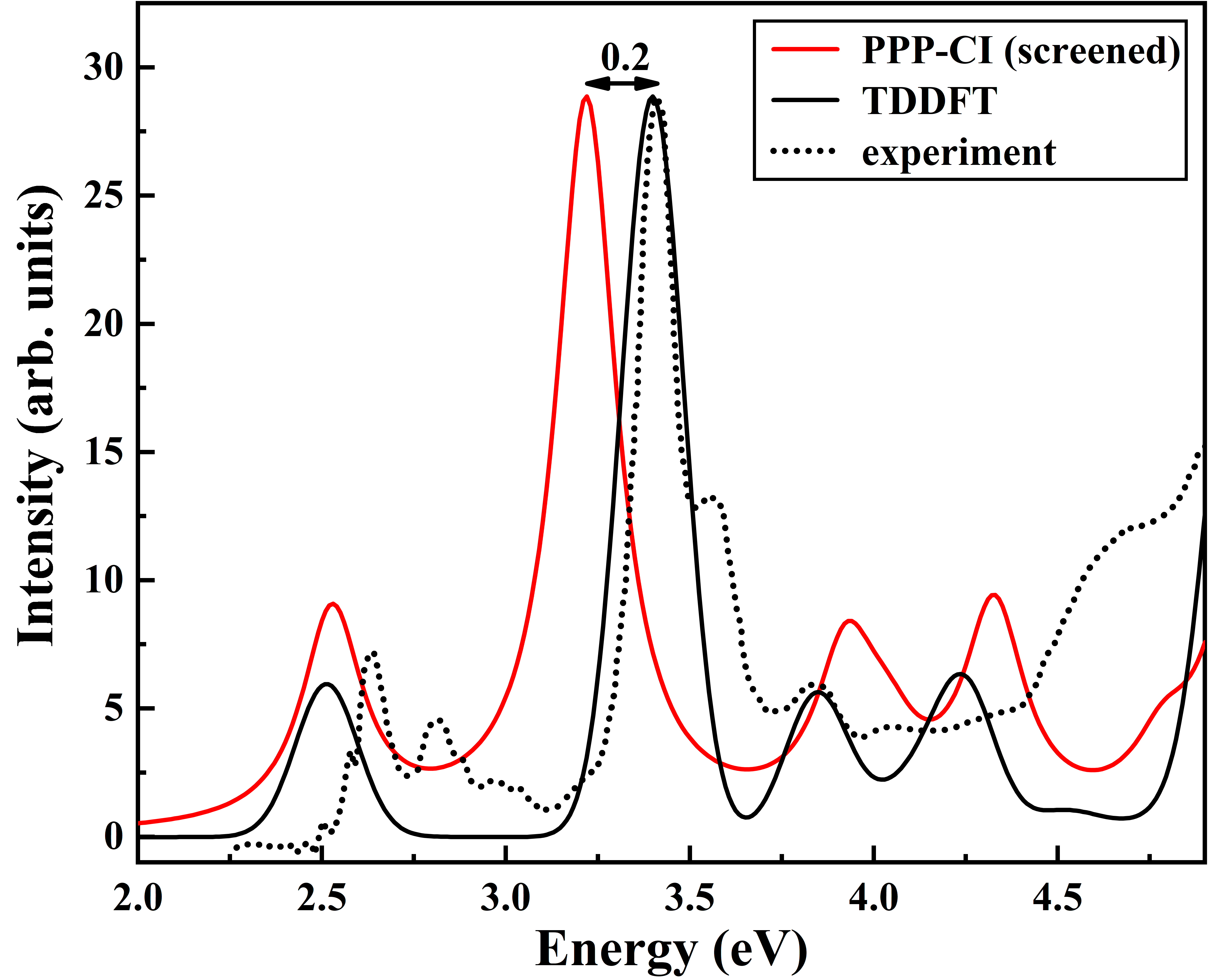}
\par\end{centering}
\caption{The comparison of the linear optical absorption spectra of GQD-36-$C_{2h}$
(dinaphtho{[}8,1,2abc;2´,1´,8´jkl{]}coronene) computed using the TDDFT
(solid black line) and screened parameter PPP-CI (solid red line)
approaches, with the experimental spectrum (dotted line) reproduced
with permission from ref. \cite{fetzer2000large} Copyright 2001,
(John Wiley and Sons). The calculated spectra have been scaled so
that their MI peaks have the same intensity as the experimental MI
peak. We have converted the original experimental spectra reported
in the wavelength (nm) units to energy (eV) units for direct comparison
with our calculations. \protect\label{fig:Comparison-of-theory-expt-36-c2h}}

\end{figure}
\par\end{center}

The first-principles TDDFT spectrum also shows very good qualitative
and quantitative agreement with the experimental one. The TDDFT location
of the MI peak (3.40 eV) is in excellent agreement with the experimental
value 3.41 eV reported by Fetzer \cite{fetzer2000large}. Except the
most intense one, peaks at 2.51 eV, 3.84 eV are very close to the
experimental peaks at 2.50 eV, 3.85 eV respectively. 

\subsection*{Influence of symmetry }

It is also important to compare the optical properties of dinaphtho{[}8,1,2abc;2´,1´,8´klm{]}coronene
(GQD-36-$C_{2v}$) and dinaphtho{[}8,1,2abc;2´,1´,8´jkl{]}coronene
(GQD-36-$C_{2h}$) because they are isomers of each other with the
same chemical formula (C\textsubscript{36}H\textsubscript{16}) but
with different point groups $C_{2v}$, and $C_{2h}$, respectively.
A careful examination of their optical spectra shows significant differences
in the lower energy region. We note that at all levels of theory and
also experiment, the first peak of the absorption spectra representing
the optical gap, is red shifted for GQD-36-$C_{2h}$ compared to that
of GQD-36-$C_{2v}$. From Table \ref{tab:optical-gaps} we can see
the amount of shift is 0.28 eV for the TDDFT method, while it is 0.38
eV both for the PPP-CI method (screened parameters) as well the experiment\cite{fetzer2000large}.
Though the most intense peaks are almost at the same location at all
levels of theory and experiment for the two molecules (see Table \ref{tab:gqd-mi}),
however, this significant shift in the optical gap indicates strong
influence of symmetry on the optical properties. Furthermore, this
red-shift which has already been observed in the experimental result
reported by Fetzer\cite{fetzer2000large}, can be used for optical
detection of the two isomers.

\subsection{Singlet-triplet gap}

The singlet-triplet gap, also called the spin gap, defined as the
energy difference between the lowest singlet ($S_{0}$) and triplet
states ($T_{1}$) of a given molecule, contains important information
about a possible magnetic character as well as electron-correlation
effects in the system. In our case, it is defined as $\Delta E_{ST}=E(1^{3}B_{x})-E(1^{1}A_{y})$,
here $(x=2$, $y=1)$ for the first two molecules with the $C_{2v}$
point group symmetry, and $(x=u,$ $y=g)$ for the third molecule
with $C_{2h}$ point group symmetry. We saw in the previous section
that the many particle wave function of the first singlet state ($S_{1}$)
optically connected to $S_{0}$ is dominated by the single excitation
$|H\rightarrow L\rangle$. Similarly, the many-particle wave function
of $T_{1}$ is also dominated by the same orbital excitation $|H\rightarrow L\rangle$,
as a result of which: (a) the many-particle wave function of the $T_{1}$
state will have the same point-group symmetry of $B_{2}$($C_{2v})$
or $B_{u}$($C_{2h}$) as that of the corresponding $S_{1}$ state,
and (b) in the non-interacting tight-binding theory $T_{1}$ and $S_{1}$
will be degenerate. However, once the electron-electron interactions
are taken into account as in the PPP model, the $S_{1}$-$T_{1}$
degeneracy gets lifted leading to a spin gap significantly smaller
than the optical gap. Thus, it is obvious that the different spin
and optical gaps of a system are a consequence of electron-correlation
effects. In Table \ref{tab:Singlet-triplet-gap}, spin gaps $\Delta_{ST}$
of the three molecules calculated using PPP-CI and DFT approaches
are presented, from which we conclude: (a) screened parameter based
spin gaps are smaller than the standard parameter ones, while the
DFT spin gaps are between the standard and screened parameter values,
and (b) spin gap of each molecule is significantly smaller than its
optical gap (see Table \ref{tab:optical-gaps}) for both PPP-CI as
well as TDDFT calculations. On comparing the spin gaps of the two
isomers, similar to their optical gaps, we find the spin gap of GQD-36-$C_{2v}$
to be larger than that of GQD-36-$C_{2h}$. Given the fact that our
PPP-CI (screened) as well as TDDFT results on the optical gaps and
several higher energy features in the absorption spectra of the three
GQDs are found to be in very good agreement with the experiments,
we expect the same will hold true for spin gaps as well. 

\begin{table}
\caption{Singlet-triplet gap, or the spin gap ($\Delta_{ST}$) of various GQDs
computed using the PPP-MRSDCI approach (using the standard and screened
parameters) and using DFT. \protect\label{tab:Singlet-triplet-gap}}

\begin{tabular}{cccc}
\toprule 
GQD  & \multicolumn{2}{c}{$\Delta_{ST}$ with PPP-CI (eV)} & $\Delta_{ST}$ with DFT (eV)\tabularnewline
\cmidrule{2-4}
 & Standard & Screened & \tabularnewline
\midrule
\midrule 
GQD-30-$C_{2v}$ & 1.52 & 1.30 & 1.35\tabularnewline
GQD-36-$C_{2v}$ & 2.20 & 1.85 & 2.17\tabularnewline
GQD-36-$C_{2h}$ & 1.84 & 1.53 & 1.76\tabularnewline
\bottomrule
\end{tabular}

\end{table}

\section{Conclusions}

\label{sec:Conclusions}

In this paper we presented a systematic, large-scale computational
study of the electronic structure and optical properties of three
hydrogen-passivated GQDs of low point-group symmetries, i.e., $C_{2v}$
and $C_{2h}$. These GQDs are also known as coronene derivatives,
namely, dibenzo{[}bc,ef{]}coronene (C\textsubscript{30}H\textsubscript{14}),
and two isomers, dinaphtho{[}8,1,2abc;2´,1´,8´klm{]}coronene (C\textsubscript{36}H\textsubscript{16})
and dinaphtho{[}8,1,2abc; 2´,1´,8´jkl{]}coronene (C\textsubscript{36}H\textsubscript{16}).
As far as the electronic properties are concerned, we presented and
discussed their single-particle energy levels for the orbitals close
to the Fermi level using both the DFT and PPP methodologies. As expected,
the HOMO-LUMO and other energy gaps computed using the DFT approach
were found to be lower compared to the corresponding values calculated
with PPP-Hartree-Fock methodology. The analysis of the molecular orbitals
and the density of states indicates that the optical absorption spectra
of these GQDs are mainly dominated by $\pi\rightarrow\pi^{*}$ excitations,
with negligible contributions from $\sigma/\sigma*$ orbitals, in
full agreement with the $\sigma-\pi$ separation assumptions for the
planar PAHs, at the heart of the PPP model. For the purpose of studying
optical properties, we employed both the TDDFT approach, as well as
PPP model based approach. For the PPP model Hamiltonian based calculations,
in order to include the electron-correlation effects beyond the mean-field
Hartree-Fock calculations, we employed the MRSDCI approach using both
the standard and screened parameters. Optical spectra obtained with
screened parameters are red shifted as compared to the spectra obtained
with standard parameters for all the three molecules. The computed
optical gaps by the PPP-CI approach coupled with the screened parameters
were found to be in excellent agreement with the experimental results.
However, TDDFT calculations were found to be very accurate in predicting
the locations of both the optical gaps as well as most-intense peaks
in the spectra of all the three GQDs, when compared to the experiments.
In both the TDDFT and the PPP-CI spectra computed with the screened
parameters, we found that several other peaks and features besides
the first and MI peaks were close to the experimentally measured values
for both the isomers of GQD-36. Additionally, with the PPP-CI methodology
and the DFT approach we also computed the spin gaps of all the three
GQDs, which are significantly lower than their corresponding optical
gaps. Comparison shows values of spin gaps calculated with screened
parameters are lower compared to the corresponding values calculated
with standard parameters within the PPP-CI approach, while the DFT
values are in between those. These calculated spin gaps can be tested
in the future experiments.

It will also be useful to study the nonlinear optical response of
these systems such as the two-photon absorption and third-harmonic
generation. We plan to undertake those calculations in the future
along with the influence of doping these systems with heteroatoms. 

\bibliographystyle{achemso}
\bibliography{low_sym_ref}

\end{document}


\title{Supporting Information:\\
DFT and Model Hamiltonian Study of Optoelectronic Properties of Some
Low-symmetry Graphene Quantum Dots}
\author{Samayita Das}
\email{samayitadas5@gmail.com}

\author{Alok Shukla}
\email{shukla@iitb.ac.in}

\address{Department of Physics, Indian Institute of Technology Bombay, Powai,
Mumbai 400076, India`}

\maketitle
In this supporting information, the following tables show the information
about the excited states contributing to important peaks in the calculated
absorption spectra of the lower-symmetry graphene quantum dots considered
in this work. The excited state symmetries of the corresponding peaks
are also presented below. Here, $f$ (= $\frac{2\times m_{e}}{3\times\hbar^{2}}(E)\sum_{j=x,y,z}|\langle e|O_{j}|r\rangle|^{2}$)
indicates the oscillator strength for a particular transition where
$E$, $|e\rangle$, $|r\rangle$ and $O_{j}$ representing the peak
energy, excited states of the corresponding peak, reference state
and electric dipole operator for different Cartesian components respectively.
In the last column, the numbers inside the bracket represent the contributing
coefficients of the wave functions for each corresponding configuration.

\subsubsection*{Data from TDDFT calculations}

Detailed information related to the UV-vis optical absorption spectra
of the three GQDs calculated using the TDDFT approach. This includes
peak positions $E$, symmetries of the excited states, oscillator
strengths $f$, polarization direction of the absorbed photon, and
the many-body wave functions of the excited states.

\begin{table}[H]
\caption{Information regarding the excited states contributing to the peaks
in the linear optical absorption spectrum of GQD-30-$C_{2v}$ (dibenzo{[}bc,ef{]}coronene,
C\protect\textsubscript{30}H\protect\textsubscript{14}) computed
using the first-principles TDDFT method (Fig. 3, main article). In
the 'Peak' column, subscripts $x$ ($y$) indicate polarization direction
of the absorbed photon is along $x$ ($y$) direction, while $xy$
denotes peaks with contribution from both $x$ and $y$ polarized
photon(s).}

\begin{tabular}{ccccccccc}
\hline 
Peak & \hspace{1.7cm} & Symmetry & \hspace{1.7cm} & E (eV) & \hspace{1.7cm} & $f$ & \hspace{1.7cm} & Wave function\tabularnewline
\hline 
\hline 
I\textsubscript{$x$} &  & \textsuperscript{1}B\textsubscript{2} &  & 2.22 &  & 0.23 &  & $\arrowvert H\rightarrow L\rangle(0.70260)$\tabularnewline
II\textsubscript{$y$} &  & \textsuperscript{1}A\textsubscript{1} &  & 3.63 &  & 0.21 &  & $\arrowvert H-3\rightarrow L\rangle(0.5263)$\tabularnewline
III\textsubscript{$y$} &  & \textsuperscript{1}A\textsubscript{1} &  & 3.83 &  & 0.54 &  & $\arrowvert H\rightarrow L+3\rangle(0.39140)$\tabularnewline
IV\textsubscript{$x$} &  & \textsuperscript{1}B\textsubscript{2} &  & 4.45 &  & 0.06 &  & $\arrowvert H-1\rightarrow L+1\rangle(0.61216)$\tabularnewline
V\textsubscript{$x$} &  & \textsuperscript{1}B\textsubscript{2} &  & 4.84 &  & 0.21 &  & $\arrowvert H-2\rightarrow L+2\rangle(0.50254)$\tabularnewline
 &  & \textsuperscript{1}B\textsubscript{2} &  & 4.90 &  & 0.05 &  & $\arrowvert H-2\rightarrow L+4\rangle(0.45413)$\tabularnewline
VI\textsubscript{$xy$} &  & \textsuperscript{1}B\textsubscript{2} &  & 5.30 &  & 0.16 &  & $\arrowvert H-2\rightarrow L+5\rangle(0.47175)$\tabularnewline
 &  & \textsuperscript{1}A\textsubscript{1} &  & 5.34 &  & 0.12 &  & $\arrowvert H-4\rightarrow L+3\rangle(0.21316)$\tabularnewline
\hline 
\end{tabular}
\end{table}

\begin{table}[H]
\caption{Information regarding the excited states contributing to the peaks
in the linear optical absorption spectrum of GQD-36-$C_{2v}$ (dinaphtho{[}8,1,2abc;2\textasciiacute ,1\textasciiacute ,8\textasciiacute klm{]}coronene,
C\protect\textsubscript{36}H\protect\textsubscript{16}) computed
using the first-principles TDDFT method (Fig. 5, main article). In
the 'Peak' column, subscripts $x$ ($y$) indicate polarization direction
of the absorbed photon is along $x$ ($y$) direction, while $xy$
denotes peaks with contribution from both $x$ and $y$ polarized
photon(s).}

\begin{tabular}{ccccccccc}
\hline 
Peak & \hspace{1.7cm} & Symmetry & \hspace{1.7cm} & E (eV) & \hspace{1.7cm} & $f$ & \hspace{1.7cm} & Wave function\tabularnewline
\hline 
\hline 
I\textsubscript{$x$} &  & \textsuperscript{1}B\textsubscript{2} &  & 2.79 &  & 0.20 &  & $\arrowvert H\rightarrow L\rangle(0.65252)$\tabularnewline
II\textsubscript{$xy$} &  & \textsuperscript{1}A\textsubscript{1} &  & 3.24 &  & 0.30 &  & $\arrowvert H\rightarrow L+1\rangle(0.50620)$\tabularnewline
 &  & \textsuperscript{1}B\textsubscript{2} &  & 3.36 &  & 0.71 &  & $\arrowvert H-1\rightarrow L+1\rangle(0.63007)$\tabularnewline
III\textsubscript{$x$} &  & \textsuperscript{1}B\textsubscript{2} &  & 3.86 &  & 0.05 &  & $\arrowvert H-3\rightarrow L\rangle(0.45941)$\tabularnewline
IV\textsubscript{$x$} &  & \textsuperscript{\textsuperscript{1}B\textsubscript{2}} &  & 4.07 &  & 0.18 &  & $\arrowvert H-2\rightarrow L+1\rangle(0.43385)$\tabularnewline
V\textsubscript{$x$} &  & \textsuperscript{\textsuperscript{1}B\textsubscript{2}} &  & 4.33 &  & 0.10 &  & $\arrowvert H-4\rightarrow L\rangle(0.47388)$\tabularnewline
VI\textsubscript{$y$} &  & \textsuperscript{1}A\textsubscript{1} &  & 4.70 &  & 0.24 &  & $\arrowvert H-4\rightarrow L+1\rangle(0.38807)$\tabularnewline
VII\textsubscript{$y$} &  & \textsuperscript{1}A\textsubscript{1} &  & 5.04 &  & 0.05 &  & $\arrowvert H-7\rightarrow L+1\rangle(0.34445)$\tabularnewline
\hline 
\end{tabular}
\end{table}

\begin{table}[H]
\caption{Information regarding the excited states contributing to the peaks
in the linear optical absorption spectrum of GQD-36-$C_{2h}$ (dinaphtho{[}8,1,2abc;2\textasciiacute ,1\textasciiacute ,8\textasciiacute jkl{]}
coronene, C\protect\textsubscript{36}H\protect\textsubscript{16})
computed using the first-principles TDDFT method (Fig. 8, main article).
In the 'Peak' column, $x$$y$ indicates the mixed polarization of
the absorbed photon. }

\begin{tabular}{ccccccccc}
\hline 
Peak & \hspace{1.7cm} & Symmetry & \hspace{1.7cm} & E (eV) & \hspace{1.7cm} & $f$ & \hspace{1.7cm} & Wave function\tabularnewline
\hline 
\hline 
I\textsubscript{$xy$} &  & \textsuperscript{1}B\textsubscript{u} &  & 2.51 &  & 0.22 &  & $\arrowvert H\rightarrow L\rangle(0.68754)$\tabularnewline
II\textsubscript{$x$y} &  & \textsuperscript{1}B\textsubscript{u} &  & 3.40 &  & 1.09 &  & $\arrowvert H-1\rightarrow L\rangle(0.49838)$\tabularnewline
III\textsubscript{$xy$} &  & \textsuperscript{1}B\textsubscript{u} &  & 3.84 &  & 0.21 &  & $\arrowvert H-1\rightarrow L+1\rangle(0.62540)$\tabularnewline
IV\textsubscript{$xy$} &  & \textsuperscript{1}B\textsubscript{u} &  & 4.24 &  & 0.23 &  & $\arrowvert H-2\rightarrow L+2\rangle(0.50781)$\tabularnewline
V\textsubscript{$xy$} &  & \textsuperscript{1}B\textsubscript{u} &  & 4.50 &  & 0.03 &  & $\arrowvert H-3\rightarrow L+2\rangle(0.49022)$\tabularnewline
VI\textsubscript{$xy$} &  & \textsuperscript{1}B\textsubscript{u} &  & 4.97 &  & 0.35 &  & $\arrowvert H-3\rightarrow L+2\rangle(0.37755)$\tabularnewline
 &  & \textsuperscript{1}B\textsubscript{u} &  & 5.01 &  & 0.47 &  & $\arrowvert H-2\rightarrow L+6\rangle(0.42446)$\tabularnewline
\hline 
\end{tabular}
\end{table}

\subsubsection*{Data from PPP model based calculations}

Detailed information related to the UV-vis optical absorption spectra
of the three GQDs calculated using the PPP-CI approach. This includes
peak positions $E$, symmetries of the excited states, oscillator
strengths $f$, polarization direction of the absorbed photon, and
the many-body wave functions of the excited states.

\begin{table}[H]
\caption{Information regarding the excited states contributing to the peaks
in the linear optical absorption spectrum of GQD-30-$C_{2v}$ (dibenzo{[}bc,ef{]}coronene,
C\protect\textsubscript{30}H\protect\textsubscript{14}) computed
using the PPP-CI method and the standard parameters (Fig. 4, main
article). The rest of the information is same as in the caption of
Table S1. }

\begin{tabular}{ccccccccc}
\hline 
Peak & \hspace{1.7cm} & Symmetry & \hspace{1.7cm} & E (eV) & \hspace{1.7cm} & $f$ & \hspace{1.7cm} & Wave function\tabularnewline
\hline 
\hline 
I\textsubscript{$x$} &  & \textsuperscript{1}B\textsubscript{2} &  & 2.76 &  & 0.39 &  & $\arrowvert H\rightarrow L\rangle(0.8528)$\tabularnewline
 &  &  &  &  &  &  &  & $\arrowvert H-1\rightarrow L+1\rangle(0.1641)$\tabularnewline
II\textsubscript{$y$} &  & \textsuperscript{1}A\textsubscript{1} &  & 4.29 &  & 1.10 &  & $\arrowvert H-1\rightarrow L\rangle+c.c.(0.5495)$\tabularnewline
III\textsubscript{$x$} &  & \textsuperscript{1}B\textsubscript{2} &  & 4.81 &  & 0.20 &  & $\arrowvert H-1\rightarrow L+1\rangle(0.5319)$\tabularnewline
 &  &  &  &  &  &  &  & $\arrowvert H-2\rightarrow L+2\rangle(0.2664)$\tabularnewline
IV\textsubscript{$x$} &  & \textsuperscript{1}B\textsubscript{2} &  & 5.46 &  & 0.46 &  & $\arrowvert H-3\rightarrow L+3\rangle(0.4181)$\tabularnewline
 &  &  &  &  &  &  &  & $\arrowvert H-2\rightarrow L+2\rangle(0.4062)$\tabularnewline
V\textsubscript{$xy$} &  & \textsuperscript{1}A\textsubscript{1} &  & 5.61 &  & 0.47 &  & $\arrowvert H-1\rightarrow L+2\rangle+c.c.(0.3499)$\tabularnewline
 &  & \textsuperscript{1}B\textsubscript{2} &  & 5.65 &  & 0.48 &  & $\arrowvert H-2\rightarrow L+2\rangle(0.3281)$\tabularnewline
 &  &  &  &  &  &  &  & {\small$\arrowvert H\rightarrow L;H\rightarrow L+1\rangle+c.c.(0.2480)$}{\scriptsize{} }\tabularnewline
VI\textsubscript{$y$} &  & \textsuperscript{1}A\textsubscript{1} &  & 5.78 &  & 0.41 &  & $\arrowvert H-4\rightarrow L+1\rangle+c.c.(0.4716)$\tabularnewline
VII\textsubscript{$xy$} &  & \textsuperscript{1}A\textsubscript{1} &  & 5.98 &  & 0.53 &  & $\arrowvert H-6\rightarrow L\rangle+c.c.(0.2649)$\tabularnewline
 &  & \textsuperscript{1}B\textsubscript{2} &  & 5.99 &  & 0.68 &  & $\arrowvert H-3\rightarrow L+3\rangle(0.4541)$\tabularnewline
 &  &  &  &  &  &  &  & $\arrowvert H-2\rightarrow L+2\rangle(0.3634)$\tabularnewline
VIII\textsubscript{$x$} &  & \textsuperscript{1}B\textsubscript{2} &  & 6.18 &  & 0.37 &  & $\arrowvert H-2\rightarrow L+4\rangle+c.c.(0.4100)$\tabularnewline
IX\textsubscript{$xy$} &  & \textsuperscript{1}A\textsubscript{1} &  & 6.45 &  & 0.44 &  & $\arrowvert H-9\rightarrow L\rangle+c.c.(0.3448)$\tabularnewline
 &  & \textsuperscript{1}B\textsubscript{2} &  & 6.46 &  & 0.19 &  & $\arrowvert H-4\rightarrow L+4\rangle(0.6639)$\tabularnewline
 &  &  &  &  &  &  &  & {\small$\arrowvert H\rightarrow L;H-4\rightarrow L+1\rangle+c.c.(0.2161)$}\tabularnewline
 &  & \textsuperscript{1}A\textsubscript{1} &  & 6.57 &  & 0.79 &  & $\arrowvert H-1\rightarrow L+5\rangle+c.c.(0.3942)$\tabularnewline
\hline 
\end{tabular}
\end{table}

\begin{table}[H]
\caption{Information regarding the excited states contributing to the peaks
in the linear optical absorption spectrum of GQD-30-$C_{2v}$ (dibenzo{[}bc,ef{]}coronene,
C\protect\textsubscript{30}H\protect\textsubscript{14}) computed
using the PPP-CI method and the screened parameters (Fig. 4, main
article). The rest of the information is same as in the caption of
Table S1. }

\begin{tabular}{ccccccccc}
\hline 
Peak & \hspace{1.7cm} & Symmetry & \hspace{1.7cm} & E (eV) & \hspace{1.7cm} & $f$ & \hspace{1.7cm} & Wave function\tabularnewline
\hline 
\hline 
I\textsubscript{$x$} &  & \textsuperscript{1}B\textsubscript{2} &  & 2.36 &  & 0.60 &  & $\arrowvert H\rightarrow L\rangle(0.8685)$\tabularnewline
 &  &  &  &  &  &  &  & $\arrowvert H-2\rightarrow L+2\rangle(0.0634)$\tabularnewline
II\textsubscript{$y$} &  & \textsuperscript{1}A\textsubscript{1} &  & 3.41 &  & 1.13 &  & $\arrowvert H-1\rightarrow L\rangle+c.c.(0.5907)$\tabularnewline
III\textsubscript{$y$} &  & \textsuperscript{1}A\textsubscript{1} &  & 3.83 &  & 0.05 &  & $\arrowvert H\rightarrow L+3\rangle+c.c.(0.5942)$\tabularnewline
IV\textsubscript{$x$} &  & \textsuperscript{1}B\textsubscript{2} &  & 4.30 &  & 0.40 &  & $\arrowvert H-1\rightarrow L+1\rangle(0.7471)$\tabularnewline
 &  &  &  &  &  &  &  & $\arrowvert H-4\rightarrow L\rangle+c.c.(0.1963)$\tabularnewline
V\textsubscript{$xy$} &  & \textsuperscript{1}B\textsubscript{2} &  & 4.69 &  & 0.77 &  & $\arrowvert H-2\rightarrow L+2\rangle(0.7999)$\tabularnewline
 &  &  &  &  &  &  &  & $\arrowvert H\rightarrow L;H\rightarrow L;H-2\rightarrow L+2\rangle(0.1390)$\tabularnewline
 &  & \textsuperscript{1}A\textsubscript{1} &  & 4.70 &  & 0.23 &  & $\arrowvert H-6\rightarrow L\rangle+c.c.(0.3107)$\tabularnewline
VI\textsubscript{$y$} &  & \textsuperscript{1}A\textsubscript{1} &  & 4.93 &  & 0.32 &  & $\arrowvert H-3\rightarrow L+2\rangle+c.c.(0.4402)$\tabularnewline
VII\textsubscript{$y$} &  & \textsuperscript{1}A\textsubscript{1} &  & 5.23 &  & 0.22 &  & $\arrowvert H-4\rightarrow L+1\rangle+c.c.(0.4897)$\tabularnewline
VIII\textsubscript{$y$} &  & \textsuperscript{1}A\textsubscript{1} &  & 5.48 &  & 0.61 &  & $\arrowvert H-1\rightarrow L+5\rangle+c.c.(0.4422)$\tabularnewline
IX\textsubscript{$x$} &  & \textsuperscript{1}B\textsubscript{2} &  & 6.20 &  & 0.03 &  & $\arrowvert H\rightarrow L;H-2\rightarrow L+1\rangle+c.c.(0.5387)$\tabularnewline
X\textsubscript{$x$} &  & \textsuperscript{1}B\textsubscript{2} &  & 6.52 &  & 0.07 &  & $\arrowvert H-4\rightarrow L+2\rangle+c.c.(0.3788)$\tabularnewline
\hline 
\end{tabular}
\end{table}

\begin{table}[H]
\caption{Information regarding the excited states contributing to the peaks
in the linear optical absorption spectrum of GQD-36-$C_{2v}$ (dinaphtho{[}8,1,2abc;2\textasciiacute ,1\textasciiacute ,8\textasciiacute klm{]}coronene,
C\protect\textsubscript{36}H\protect\textsubscript{16}) computed
using the PPP-CI method and the standard parameters (Fig. 6, main
article). The rest of the information is same as in the caption of
Table S2. }

\begin{tabular}{ccccccccc}
\hline 
Peak & \hspace{1.7cm} & Symmetry & \hspace{1.7cm} & E (eV) & \hspace{1.7cm} & $f$ & \hspace{1.7cm} & Wave function\tabularnewline
\hline 
\hline 
I\textsubscript{$x$} &  & \textsuperscript{1}B\textsubscript{2} &  & 3.22 &  & 0.08 &  & $\arrowvert H\rightarrow L\rangle(0.6964)$\tabularnewline
 &  &  &  &  &  &  &  & $\arrowvert H-1\rightarrow L+1\rangle(0.4872)$\tabularnewline
II\textsubscript{$xy$} &  & \textsuperscript{1}B\textsubscript{2} &  & 3.93 &  & 0.96 &  & $\arrowvert H-1\rightarrow L+1\rangle(0.6650)$\tabularnewline
 &  &  &  &  &  &  &  & $\arrowvert H\rightarrow L\rangle(0.4766)$\tabularnewline
 &  & \textsuperscript{1}A\textsubscript{1} &  & 3.99 &  & 0.59 &  & $\arrowvert H-1\rightarrow L\rangle+c.c.(0.4348)$\tabularnewline
III\textsubscript{$y$} &  & \textsuperscript{1}A\textsubscript{1} &  & 4.31 &  & 0.35 &  & $\arrowvert H-1\rightarrow L\rangle+c.c.(0.3905)$\tabularnewline
IV\textsubscript{$x$} &  & \textsuperscript{1}B\textsubscript{2} &  & 4.59 &  & 0.05 &  & $\arrowvert H-5\rightarrow L+1\rangle+c.c.(0.4965)$\tabularnewline
V\textsubscript{$y$} &  & \textsuperscript{1}A\textsubscript{1} &  & 4.95 &  & 0.46 &  & $\arrowvert H-5\rightarrow L\rangle+c.c.(0.3600)$\tabularnewline
VI\textsubscript{$x$} &  & \textsuperscript{1}B\textsubscript{2} &  & 5.07 &  & 0.27 &  & $\arrowvert H-1\rightarrow L+2\rangle+c.c.(0.4051)$\tabularnewline
VII\textsubscript{$y$} &  & \textsuperscript{1}A\textsubscript{1} &  & 5.41 &  & 0.14 &  & $\arrowvert H-4\rightarrow L+1\rangle+c.c.(0.5341)$\tabularnewline
VIII\textsubscript{$y$} &  & \textsuperscript{1}A\textsubscript{1} &  & 5.77 &  & 0.21 &  & $\arrowvert H-5\rightarrow L\rangle+c.c.(0.4114)$\tabularnewline
IX\textsubscript{$x$} &  & \textsuperscript{1}B\textsubscript{2} &  & 6.00 &  & 0.70 &  & $\arrowvert H-4\rightarrow L+4\rangle(0.3526)$\tabularnewline
 &  &  &  &  &  &  &  & $\arrowvert H-6\rightarrow L+1\rangle+c.c.(0.3099)$\tabularnewline
X\textsubscript{$xy$} &  & \textsuperscript{1}B\textsubscript{2} &  & 6.32 &  & 0.48 &  & $\arrowvert H-3\rightarrow L+3\rangle(0.4172)$\tabularnewline
 &  &  &  &  &  &  &  & $\arrowvert H-4\rightarrow L+4\rangle(0.3191)$\tabularnewline
 &  & \textsuperscript{1}B\textsubscript{2} &  & 6.41 &  & 0.47 &  & $\arrowvert H-2\rightarrow L+5\rangle+c.c.(0.3571)$\tabularnewline
 &  & \textsuperscript{1}A\textsubscript{1} &  & 6.49 &  & 0.90 &  & $\arrowvert H-2\rightarrow L+3\rangle+c.c.(0.4692)$\tabularnewline
XI\textsubscript{$x$} &  & \textsuperscript{1}B\textsubscript{2} &  & 6.74 &  & 0.26 &  & $\arrowvert H-4\rightarrow L+4\rangle(0.3329)$\tabularnewline
 &  &  &  &  &  &  &  & $\arrowvert H\rightarrow L+11\rangle+c.c.(0.2803)$\tabularnewline
\hline 
\end{tabular}
\end{table}

\begin{table}[H]
\caption{Information regarding the excited states contributing to the peaks
in the linear optical absorption spectrum of GQD-36-$C_{2v}$ (dinaphtho{[}8,1,2abc;2\textasciiacute ,1\textasciiacute ,8\textasciiacute klm{]}coronene,
C\protect\textsubscript{36}H\protect\textsubscript{16}) computed
using the PPP-CI method and the screened parameters (Fig. 6, main
article). The rest of the information is same as in the caption of
Table S2. }

\begin{tabular}{ccccccccc}
\hline 
Peak & \hspace{1.7cm} & Symmetry & \hspace{1.7cm} & E (eV) & \hspace{1.7cm} & $f$ & \hspace{1.7cm} & Wave function\tabularnewline
\hline 
\hline 
I\textsubscript{$x$} &  & \textsuperscript{1}B\textsubscript{2} &  & 2.91 &  & 0.86 &  & $\arrowvert H\rightarrow L\rangle(0.8599)$\tabularnewline
 &  &  &  &  &  &  &  & $\arrowvert H-1\rightarrow L+1\rangle(0.0520)$\tabularnewline
II\textsubscript{$xy$} &  & \textsuperscript{1}A\textsubscript{1} &  & 3.17 &  & 1.09 &  & $\arrowvert H\rightarrow L+1\rangle+c.c.(0.6011)$\tabularnewline
 &  & \textsuperscript{1}B\textsubscript{2} &  & 3.19 &  & 1.06 &  & $\arrowvert H-1\rightarrow L+1\rangle(0.8517)$\tabularnewline
 &  &  &  &  &  &  &  & $\arrowvert H-1\rightarrow L+2\rangle+c.c.(0.0659)$\tabularnewline
III\textsubscript{$xy$} &  & \textsuperscript{1}A\textsubscript{1} &  & 3.92 &  & 0.04 &  & $\arrowvert H-2\rightarrow L\rangle+c.c.(0.5742)$\tabularnewline
 &  & \textsuperscript{1}B\textsubscript{2} &  & 4.00 &  & 0.31 &  & $\arrowvert H-2\rightarrow L+1\rangle+c.c.(0.5288)$\tabularnewline
IV\textsubscript{$y$} &  & \textsuperscript{1}A\textsubscript{1} &  & 4.22 &  & 0.12 &  & $\arrowvert H-1\rightarrow L+3\rangle+c.c.(0.5328)$\tabularnewline
V\textsubscript{$xy$} &  & \textsuperscript{1}B\textsubscript{2} &  & 4.40 &  & 0.16 &  & $\arrowvert H\rightarrow L+4\rangle+c.c.(0.5655)$\tabularnewline
 &  & \textsuperscript{1}A\textsubscript{1} &  & 4.46 &  & 0.27 &  & $\arrowvert H-4\rightarrow L+1\rangle+c.c.(0.4720)$\tabularnewline
VI\textsubscript{$y$} &  & \textsuperscript{1}A\textsubscript{1} &  & 5.07 &  & 0.70 &  & $\arrowvert H-3\rightarrow L+2\rangle+c.c.(0.5427)$\tabularnewline
VII\textsubscript{$x$} &  & \textsuperscript{1}B\textsubscript{2} &  & 5.33 &  & 0.72 &  & $\arrowvert H-4\rightarrow L+3\rangle+c.c.(0.4021)$\tabularnewline
 &  & \textsuperscript{1}B\textsubscript{2} &  & 5.38 &  & 0.61 &  & $\arrowvert H-1\rightarrow L+8\rangle+c.c.(0.3833)$\tabularnewline
VIII\textsubscript{$x$} &  & \textsuperscript{1}B\textsubscript{2} &  & 5.66 &  & 0.08 &  & $\arrowvert H-4\rightarrow L+4\rangle(0.7716)$\tabularnewline
 &  &  &  &  &  &  &  & $\arrowvert H-4\rightarrow L+3\rangle+c.c.(0.2137)$\tabularnewline
IX\textsubscript{$y$} &  & \textsuperscript{1}A\textsubscript{1} &  & 6.07 &  & 0.03 &  & $\arrowvert H\rightarrow L;H\rightarrow L+3\rangle+c.c.(0.5153)$\tabularnewline
X\textsubscript{$y$} &  & \textsuperscript{1}A\textsubscript{1} &  & 6.35 &  & 0.04 &  & $\arrowvert H-6\rightarrow L\rangle+c.c.(0.4991)$\tabularnewline
\hline 
\end{tabular}
\end{table}

\begin{table}[H]
\caption{Information regarding the excited states contributing to the peaks
in the linear optical absorption spectrum of GQD-36-$C_{2h}$ (dinaphtho{[}8,1,2abc;2\textasciiacute ,1\textasciiacute ,8\textasciiacute jkl{]}coronene,
C\protect\textsubscript{36}H\protect\textsubscript{16}) computed
using the PPP-CI method and the standard parameters (Fig. 9, main
article). The rest of the information is same as in the caption of
Table S3. }

\begin{tabular}{ccccccccc}
\hline 
Peak & \hspace{1.7cm} & Symmetry & \hspace{1.7cm} & E (eV) & \hspace{1.7cm} & $f$ & \hspace{1.7cm} & Wave function\tabularnewline
\hline 
\hline 
I\textsubscript{$xy$} &  & \textsuperscript{1}B\textsubscript{u} &  & 2.97 &  & 0.41 &  & $\arrowvert H\rightarrow L\rangle(0.8163)$\tabularnewline
 &  &  &  &  &  &  &  & $\arrowvert H-1\rightarrow L+1\rangle(0.2750)$\tabularnewline
II\textsubscript{$xy$} &  & \textsuperscript{1}B\textsubscript{u} &  & 4.05 &  & 1.97 &  & $\arrowvert H\rightarrow L+1\rangle+c.c.(0.5634)$\tabularnewline
III\textsubscript{$xy$} &  & \textsuperscript{1}B\textsubscript{u} &  & 4.37 &  & 0.47 &  & $\arrowvert H-2\rightarrow L+2\rangle(0.4453)$\tabularnewline
 &  &  &  &  &  &  &  & $\arrowvert H-1\rightarrow L+1\rangle(0.4182)$\tabularnewline
IV\textsubscript{$xy$} &  & \textsuperscript{1}B\textsubscript{u} &  & 4.66 &  & 0.62 &  & $\arrowvert H-1\rightarrow L+1\rangle(0.5869)$\tabularnewline
 &  &  &  &  &  &  &  & $\arrowvert H-5\rightarrow L\rangle+c.c.(0.3604)$\tabularnewline
V\textsubscript{$xy$} &  & \textsuperscript{1}B\textsubscript{u} &  & 5.34 &  & 0.31 &  & $\arrowvert H\rightarrow L+8\rangle+c.c.(0.3657)$\tabularnewline
VI\textsubscript{$xy$} &  & \textsuperscript{1}B\textsubscript{u} &  & 5.48 &  & 0.64 &  & $\arrowvert H-2\rightarrow L+2\rangle(0.5942)$\tabularnewline
 &  &  &  &  &  &  &  & $\arrowvert H-1\rightarrow L+1\rangle(0.3272)$\tabularnewline
VII\textsubscript{$xy$} &  & \textsuperscript{1}B\textsubscript{u} &  & 5.82 &  & 0.28 &  & $\arrowvert H-7\rightarrow L+2\rangle+c.c.(0.2774)$\tabularnewline
VIII\textsubscript{$xy$} &  & \textsuperscript{1}B\textsubscript{u} &  & 6.10 &  & 0.65 &  & $\arrowvert H-1\rightarrow L+5\rangle+c.c.(0.3209)$\tabularnewline
IX\textsubscript{$xy$} &  & \textsuperscript{1}B\textsubscript{u} &  & 6.47 &  & 0.67 &  & $\arrowvert H-4\rightarrow L+4\rangle(0.5862)$\tabularnewline
 &  &  &  &  &  &  &  & $\arrowvert H-8\rightarrow L+1\rangle+c.c.(0.2342)$\tabularnewline
 &  & \textsuperscript{1}B\textsubscript{u} &  & 6.51 &  & 0.71 &  & $\arrowvert H-1\rightarrow L+5\rangle+c.c.(0.3351)$\tabularnewline
\hline 
\end{tabular}
\end{table}

\begin{table}[H]
\caption{Information regarding the excited states contributing to the peaks
in the linear optical absorption spectrum of GQD-36-$C_{2h}$ (dinaphtho{[}8,1,2abc;2\textasciiacute ,1\textasciiacute ,8\textasciiacute jkl{]}coronene,
C\protect\textsubscript{36}H\protect\textsubscript{16}) computed
using the PPP-CI method and the screened parameters (Fig. 9, main
article). The rest of the information is same as in the caption of
Table S3. }

\begin{tabular}{ccccccccc}
\hline 
Peak & \hspace{1.7cm} & Symmetry & \hspace{1.7cm} & E (eV) & \hspace{1.7cm} & $f$ & \hspace{1.7cm} & Wave function\tabularnewline
\hline 
\hline 
I\textsubscript{$xy$} &  & \textsuperscript{1}B\textsubscript{u} &  & 2.53 &  & 0.62 &  & $\arrowvert H\rightarrow L\rangle(0.8661)$\tabularnewline
 &  &  &  &  &  &  &  & $\arrowvert H-1\rightarrow L+1\rangle(0.0735)$\tabularnewline
II\textsubscript{$xy$} &  & \textsuperscript{1}B\textsubscript{u} &  & 3.22 &  & 1.98 &  & $\arrowvert H\rightarrow L+1\rangle+c.c.(0.6136)$\tabularnewline
III\textsubscript{$xy$} &  & \textsuperscript{1}B\textsubscript{u} &  & 3.93 &  & 0.41 &  & $\arrowvert H-1\rightarrow L+1\rangle(0.8552)$\tabularnewline
 &  &  &  &  &  &  &  & $\arrowvert H-2\rightarrow L+2\rangle(0.1195)$\tabularnewline
IV\textsubscript{$xy$} &  & \textsuperscript{1}B\textsubscript{u} &  & 4.33 &  & 0.67 &  & $\arrowvert H-2\rightarrow L+2\rangle(0.7508)$\tabularnewline
 &  &  &  &  &  &  &  & $\arrowvert H-5\rightarrow L\rangle+c.c.(0.2902)$\tabularnewline
V\textsubscript{$x$$y$} &  & \textsuperscript{1}B\textsubscript{u} &  & 4.79 &  & 0.21 &  & $\arrowvert H-4\rightarrow L+2\rangle+c.c.(0.3566)$\tabularnewline
VI\textsubscript{$xy$} &  & \textsuperscript{1}B\textsubscript{u} &  & 5.03 &  & 0.69 &  & $\arrowvert H-5\rightarrow L+1\rangle+c.c.(0.3672)$\tabularnewline
VII\textsubscript{$xy$} &  & \textsuperscript{1}B\textsubscript{u} &  & 5.75 &  & 0.07 &  & $\arrowvert H-7\rightarrow L+2\rangle+c.c.(0.5930)$\tabularnewline
\hline 
\end{tabular}
\end{table}

\subsection*{Density of States}
\begin{flushleft}
\begin{figure}[H]
\begin{centering}
\includegraphics[width=15cm]{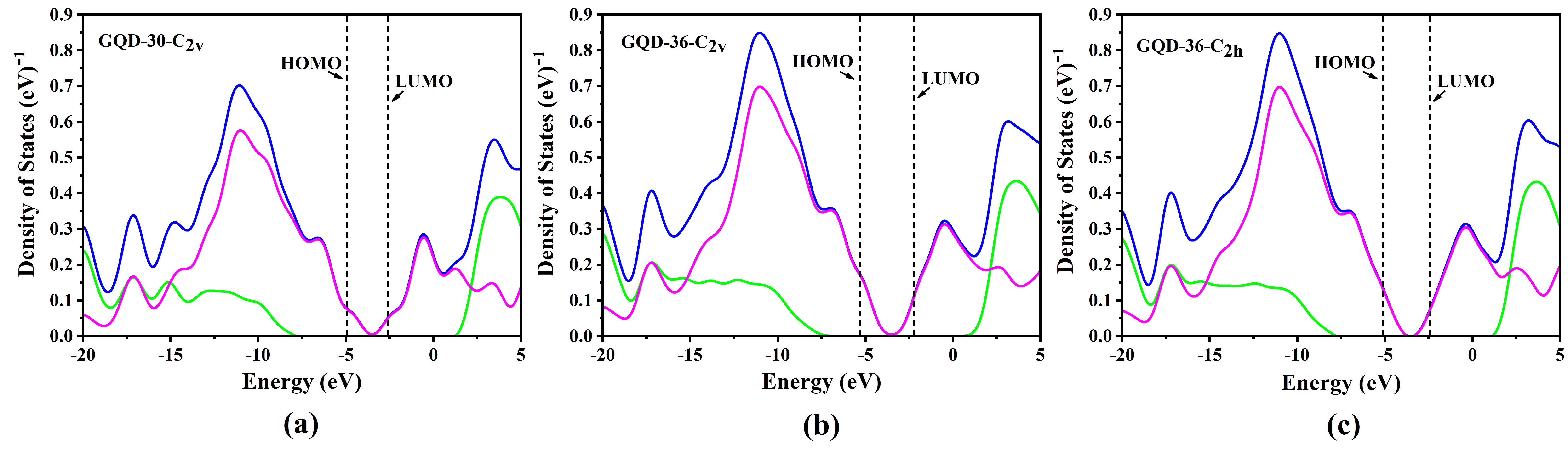}
\par\end{centering}
\caption{The total and orbital-projected density of states of the three H-passivated
GQDs: (a) GQD-30-C\protect\textsubscript{2v}, (b) GQD-36-C\protect\textsubscript{2v},
(c) GQD-36-C\protect\textsubscript{2h}. Blue lines denote the total
density of states (TDOS), while magenta and green lines represent
the partial DOS corresponding to the contributions of the s and p
orbitals, respectively. It is noteworthy that the DOS near HOMO/LUMO
is completely dominated by the contribution of the p orbitals, consistent
with the $\pi/\pi^{*}$ nature of the states in that region. \protect\label{fig:pDOS}}
\end{figure}
\par\end{flushleft}

\subsection*{The influence of the exchange-correlation functional on the TDDFT
optical absorption spectra }
\begin{center}
\begin{table}[H]
\caption{\protect\label{tab:gqd-mi}The optical gaps and positions of the most
intense (MI) peaks in the optical absorption spectra of the smallest
PAH considered in this work, i.e., dibenzo{[}bc,ef{]}coronene (C\protect\textsubscript{30}H\protect\textsubscript{14},
denoted as GQD-30-$C_{2v}$), computed using three different functionals,
and the 6-31++G(d,p) basis set.}

\centering{}%
\begin{tabular}{>{\centering}p{3cm}cc}
\toprule 
Different & Optical gap & Most intense peak\tabularnewline
\midrule 
Functionals & (eV) & (eV)\tabularnewline
\midrule
\midrule 
B3LYP & 2.22 & 3.83\tabularnewline
HSE06 & 2.23 & 3.88\tabularnewline
PBE & 1.99 & 3.47\tabularnewline
\bottomrule
\end{tabular}
\end{table}
\par\end{center}